\documentclass[aps,prc,reprint,groupedaddress,superscriptaddress]{revtex4-1}  % v2
\usepackage[utf8]{inputenc}
\usepackage{amssymb,amsmath,amsfonts}
\usepackage{multirow}
\usepackage{slashed} 
\usepackage{graphicx}
\usepackage{subfigure}
\usepackage{hyperref}
\usepackage{float}

\begin{document}
  \bibliographystyle{apsrev4-1}
  \title{Light mesons within the basis light-front quantization framework}
  \author{Wenyang Qian}
  \email{wqian@iastate.edu}
  \affiliation{Department of Physics and Astronomy, Iowa State University, Ames, IA, 50011, USA}
  \author{Shaoyang Jia}
  \email{sjia@iastate.edu}
  \affiliation{Department of Physics and Astronomy, Iowa State University, Ames, IA, 50011, USA}
  \author{Yang Li}
  \email{leeyoung@iastate.edu}
  \affiliation{Department of Physics and Astronomy, Iowa State University, Ames, IA, 50011, USA}
  \affiliation{School of Nuclear Science and Technology, University of Chinese Academy of Sciences, Beijing 100049, China}
  \author{James P. Vary}
  \email{jvary@iastate.edu}
  \affiliation{Department of Physics and Astronomy, Iowa State University, Ames, IA, 50011, USA}
  \date{\today}
  \begin{abstract}
  We study the light-unflavored mesons as relativistic bound states in the nonperturbative Hamiltonian formalism of the basis light-front quantization (BLFQ) approach. The dynamics for the valence quarks of these mesons is specified by an effective Hamiltonian containing the one-gluon exchange interaction and the confining potentials both introduced in our previous work on heavy quarkonia, supplemented additionally by a pseudoscalar contact interaction. We diagonalize this Hamiltonian in our basis function representation to obtain the mass spectrum and the light-front wave functions (LFWFs). Based on these LFWFs, we then study the structure of these mesons by computing the electromagnetic form factors, the decay constants, the parton distribution amplitudes (PDAs), and the parton distribution functions (PDFs). Our results are comparable to those from experiments and other theoretical models.
  \end{abstract}
  \maketitle

%%%%%%%%%%%%%%% SECTION 1 %%%%%%%%%%%%%%%%
\section{Introduction}
Light mesons are important bound states of the strong interaction, as an \textit{ab initio} explanation of their structures requires nonperturbative solutions of quantum chromodynamics (QCD). Various theoretical methods exist for solving the structures of light mesons~\cite{Maris, DSE_Chang, Fischer:2014xha, Williams:2015cvx, Braun:pion_moments, Braun_v2, Arthur:light_meson, Sufian, Cloet_lattice, Segovia:2013, holography, Ahmady:dynamical_spin, Choi:pion_rho, Choi:2015, Choi:1997, Spence:1993, Crater:2002}. Among these methods based on equal time quantization, the Bethe-Salpeter equation (BSE) has been successful in describing properties of two-body bound states in a relativistic and covariant formalism~\cite{Maris, DSE_Chang, Fischer:2014xha, Williams:2015cvx}. Lattice QCD is capable of producing high-precision results for hadron spectroscopy and many other observables in the Euclidean space~\cite{Braun:pion_moments, Braun_v2, Arthur:light_meson, Sufian, Cloet_lattice, Segovia:2013}. Furthermore, inspired by the gauge/string duality, holographic QCD models~~\cite{holography, Ahmady:dynamical_spin} have been proposed to provide an interpretation of hadron structures using light-front quantization complementary to equal-time quantization approaches. 

In addition to these well-established approaches, basis light-front quantization (BLFQ) is an emerging framework to describe hadronic structures based on many-body methods using Hamiltonian dynamics that is fully relativistic and nonperturbative~\cite{vary_1stBLFQ}. Specifically, the light-front Schr\"{o}dinger equation formulated as an eigenvalue problem generates the mass spectrum as the eigenvalues and the light-front wave functions (LFWFs) as the eigenfunctions. The advantage of this approach is that calculating physical observables from LFWFs is straightforward. For instance, the electromagnetic form factor in light-front dynamics is simply the overlap of light-front wave functions. Furthermore, the same wave functions grant access to decay constants, parton distribution amplitudes (PDAs), and parton distribution functions (PDFs). Progress has been made in the applications of BLFQ to the positronium system~\cite{positronium}, heavy mesons~\cite{Li:2015zda, Li:2017mlw, Yang_frame, Li:2018uif, Lekha, Chen:VM, Chen:2018vdw, Tang_bc}, light mesons~\cite{Jia, Lan}, and many other systems~\cite{Vary_2016_proceeding, Vary_2018_proceeding}.

Our model stands apart from previous calculations by applying a light-front-Hamiltonian formalism in a basis function representation that provides access to the full spectroscopy. In this work, we naturally extend the effective Hamiltonian used for heavy quarkonia to light-unflavored mesons with the addition of a pseudoscalar interaction inspired by the Nambu--Jona-Lasinio (NJL) model~\cite{Klimt:1989pm,Vogl:1989ea,Vogl:1991qt}. The resulting Hamiltonian consists of the light-front kinetic energy, the combined transverse and the longitudinal confinement potential, the one-gluon exchange, and the pseudoscalar interaction. The Hamiltonian is then diagonalized within the valence Fock sector in a basis function representation. It should be noted that this model is different from the earlier BLFQ-NJL model~\cite{Jia} in that the mass spectroscopy that we present here is dominated by gluon dynamics. We determine the model parameters by fitting the mass of $\rho$(770), $\rho'$(1450), and $\pi(140)$. Upon solving for the remaining states of the mass spectrum, we evaluate various physical properties for the lowest mass states with available LFWFs.

We organize this paper as follows. In Sect.~II, we introduce the effective Hamiltonian and the basis space used to solve the light-front-time-independent Schr\"{o}dinger equation in the valence Fock sector of light-unflavored mesons. In Sect.~III, we present and discuss numerical results obtained with our model, including the full spectroscopy, decay constants, charge radii, magnetic moments, and quadrupole moments. In Sect.~IV, we present the PDFs and PDAs for the ground states of the pion and the rho meson, and compare the pion results with the available experiments after QCD evolution. In Sect.~V, we summarize the paper and discuss possible future developments.

%%%%%%%%%%%%%%% SECTION 2 %%%%%%%%%%%%%%%%
\section{Effective Hamiltonian and Basis Function Representation}
\subsection{The Hamiltonian}

With the aim to approach a unified study on a selection of mesons, we inherit the effective Hamiltonian that was first proposed for the heavy quarkonia~\cite{Li:2015zda, Li:2017mlw} and later also applied to mesons with unequal masses for the valence quarks~\cite{Tang_bc}. This effective Hamiltonian in the valence Fock sector is partially based on light-front holography~\cite{holography}. The longitudinal confining potential was first introduced in Ref.~\cite{Li:2015zda} to complement the traverse holographic confining potential. The one-gluon exchange interaction, derived from the leading-order effective Hamiltonian approach, was included to produce short-range high-momentum physics as well as the spin-dependent interaction of the meson structure~\cite{Li:2015zda, Li:2017mlw}. This effective Hamiltonian in a convenient but mixed representation reads
\begin{align}\label{eq:Heff}
  H_{\mathrm{eff}} &\equiv P^+P^- - \mathbf{P}^2_\perp 
  \nonumber\\
  &= \frac{\mathbf{k}_\perp^2+m_q^2}{x} +  \frac{\mathbf{k}_\perp^2+m_{\bar{q}}^2}{1-x} + \kappa^4x(1-x) \mathbf{r}_\perp^2 
  \nonumber\\
  &\quad - \frac{\kappa^4}{(m_q + m_{\bar{q}})^2}\partial_x (x(1-x)\partial_x) + V_g, 
\end{align}
where $m_q$ ($m_{\bar{q}}$) is the mass of the quark (anti-quark), $\mathbf{k}_\perp$ $(-\mathbf{k}_\perp)$ is the relative momentum of the quark (anti-quark) using total $\mathbf{P}_\perp = 0$, $x$ $(1-x)$ is the longitudinal momentum fraction of the quark (anti-quark), and $\mathbf{r}_\perp$ is the transverse separation of the quark and anti-quark. The first two terms are the kinetic energy of the quark and the anti-quark. The third and forth terms describe the confinement in both the traverse and the longitudinal directions. The former serves as the light-front anti-de Sitter/quantum chromodynamics (AdS/QCD) soft-wall potential~\cite{deTeramond:2008ht, holography}, while the later is introduced to supplement the transverse confinement to form a $3$-dimensional spherical confinement potential in the non-relativistic limit. The last term $V_g$ is the one-gluon exchange. In the momentum space, it is defined by $\langle \mathbf{k}'_{\perp}, x', s', \bar{s}'|V_g|\mathbf{k}_{\perp}, x, s, \bar{s}\rangle = -C_{\mathrm{F}} 4\pi\alpha_{\mathrm{s}}(Q^2) \bar{u}_{s'}(\mathbf{k}'_\perp, x')\gamma_\mu u_s(\mathbf{k}_\perp,x) \bar{v}_{\bar{s}}(-\mathbf{k}_\perp,1-x)\gamma^\mu $ $v_{\bar{s}'}(-\mathbf{k}'_\perp,1-x')/Q^2$, where $C_{\mathrm{F}}=(N_{\mathrm{c}}^2-1)/(2N_{\mathrm{c}})=4/3$ is the color factor with $N_c = 3$ used throughout this work, and $Q^2$ is the average 4-momentum square carried by the exchange gluon~\cite{Li:2017mlw}. The running coupling used in $V_g$ is modeled on the result of one-loop perturbative QCD (pQCD)~\cite{Li:2017mlw} and is given by,
\begin{align}\label{eq:running_coupling}
  \alpha_{\mathrm{s}}(Q^2) &= \frac{1}{\beta_0 \text{ ln}(Q^2/\Lambda^2 + \tau)} 
  \nonumber\\
  &\triangleq \frac{\alpha_{\mathrm{s}}(M_z^2)}{1+\alpha_{\mathrm{s}}(M_z^2) \beta_0 \text{ ln }(\mu^2_\text{IR} + Q^2) / (\mu^2_\text{IR} + M_z^2)},
\end{align}
where $\beta_0 = (33-2N_{\mathrm{f}})/(12\pi)$ is the QCD beta function, with $N_{\mathrm{f}}$ being the number of quark flavors, and $\alpha_{\mathrm{s}}(M_z^2) = 0.1183$ is the strong coupling at the Z-boson mass of $M_z=91.2$~GeV. The constant $\tau$ is introduced to avoid the pQCD infrared (IR) catastrophe. In practice, we use $\mu_\text{IR}=0.55$ GeV with $N_{\mathrm{f}} =3$ such that $\alpha_{\mathrm{s}}(0) = 0.89$, in accordance with the calculations of heavy quarkonia where $N_{\mathrm{f}}=4$ for charmonium and $N_{\mathrm{f}}=5$ for bottomonium~\cite{Li:2017mlw}.

The effective Hamiltonian in Eq.~\eqref{eq:Heff} has been successfully applied to solve heavy meson systems with fitted parameters $m_q$ ($m_{\bar{q}}$) and $\kappa$~\cite{Li:2015zda, Li:2017mlw, Tang_bc, Li:2018uif, Lekha}, It turns out that for the light meson system, a suitable choice of these two parameters could also produce a reasonable mass spectrum, as we will show later. However, in spite of its various similarities to heavy mesons, the light meson system has an unusual feature of the $\pi$-$\rho$ mass splitting attributed to dynamical chiral symmetry breaking. We propose to add a pseudoscalar contact interaction, 
\begin{align}\label{eq:Heff5}
  H_{\gamma_5} = \int dx^{-} \int d\mathbf{x}^{\perp} P^{+} \lambda \ \bar{\psi}(x)\gamma^5\psi(x)\bar{\psi}(x)\gamma^5\psi(x),
\end{align}
where $\psi$ ($ \bar{\psi}$) is the fermion (anti-fermion) field and $\lambda$ is the coupling strength of this interaction. We propose to include this interaction into the Hamiltonian for the following reasons. First, it is in the same spirit of the NJL model~\cite{Klimt:1989pm,Vogl:1989ea,Vogl:1991qt}, which aims to explain the effective dynamics of quarks inside the light mesons. Second, as a pseudoscalar bilinear, it mostly influences the pseudoscalar states, especially the ground state ($\pi$), and leaves the other states with total angular momentum greater than 1 almost intact. This makes it ideal to produce the $\pi$-$\rho$ splitting when added to the adopted $H_{\mathrm{eff}}$ (Eq. \eqref{eq:Heff}). Additionally, its Hamiltonian matrix element can be calculated analytically in our basis representation to be introduced in the following subsection. The complete Hamiltonian in this work takes the form of $H_{\mathrm{eff},\gamma_5} = H_{\mathrm{eff}} + H_{\gamma_5}$.

\subsection{Basis Function Representation}
With our Hamiltonian specified in the previous subsection, the mass spectrum and wave functions are obtained directly as solutions of the light-front-time-independent Schr\"{o}dinger equation 
\begin{align}\label{eq:eigenvalue_equation}
H_{\text{eff,$\gamma_5$}}|\psi_h(P,j,m_j)\rangle = M_h^2|\psi_h(P,j,m_j)\rangle,
\end{align}
where $P= (P^-, P^+, \textbf{P}_\perp)$ is the four momentum of the particle, $j$ and $m_j$ are respectively the total angular momentum and its magnetic projection for that particle. We work in the valence $|q\bar{q}\rangle$ sector such that the Fock space representation of the light meson reads: 
\begin{align}\label{eq:valence_WF}
  &|\psi_h(P, j, m_j)\rangle 
  \nonumber\\
  = &\sum_{s, \bar{s}}\int_0^1\frac{\text{d}x}{2x(1-x)}\int\frac{\text{d}^2\mathbf{k}_\perp}{(2\pi)^3}\psi_{s\bar{s}/h}^{(m_j)}(\mathbf{k}_\perp, x) 
  \nonumber \\
  &\times \frac{1}{\sqrt{N_c}}\sum_{i=1}^{N_c} b_{si}^\dagger(xP^+, \mathbf{k}_\perp+x\textbf{P}_\perp) 
  \nonumber\\
  &\times d^\dagger_{\bar{s}i}((1-x)P^+, -\mathbf{k}_\perp+(1-x)\textbf{P}_\perp)|0\rangle.
\end{align}
Here $\psi_{s\bar{s}/h}^{(m_j)}(\mathbf{k}_\perp, x)$ are the valence-sector LFWFs, where $s$ and $\bar{s}$ represent the spins of the quark and the anti-quark respectively. The LFWFs are properly normalized such that
\begin{align}\label{eq:normalization}
  \sum_{s, \bar{s}} \int_0^1 \frac{dx}{2x(1-x)} \int \frac{d^2\mathbf{k}_\perp}{(2\pi)^3}&\psi_{s \bar{s}/h'}^{(m'_j)*}(\mathbf{k}_\perp, x) \psi_{s \bar{s}/h}^{(m_j)*}(\mathbf{k}_\perp, x) 
  \nonumber\\
  &= \delta_{hh'}\delta{m_j m'_j}.
\end{align}

In solving the eigenvalue equation, we use a basis function approach, the basis light-front quantization (BLFQ), where the Hamiltonian is diagonalized within a chosen basis function representation~\cite{vary_1stBLFQ}. One major advantage of implementing BLFQ is to make numerical calculation efficient. The basis choice of this work follows that of heavy quarkonia~\cite{Li:2015zda, Li:2017mlw} which employs the eigenfunctions of the effective Hamiltonian of Eq. \eqref{eq:Heff} in the absence of the one-gluon exchange. Explicitly, we expand the LFWF into the traverse and the longitudinal basis functions with coefficients $ \psi_h(n,m,l,s,\bar{s})$:
\begin{align}\label{eq:LFWF}
\psi_{s\bar{s}/h}(\mathbf{k}_\perp,x)=\sum_{n,m,l} \psi_h(n,m,l,s,\bar{s}) \phi(\frac{\mathbf{k}_\perp}{\sqrt{x(1-x)}})\chi_l(x),
\end{align}
where the basis functions are defined as 
\begin{align}
  \phi_{nm}&(\textbf{q}_\perp) = \frac{1}{\kappa}\sqrt{\frac{4\pi n!}{(n+|m|)!}} \Big ( \frac{q_\perp}{\kappa} \Big)^{|m|} e^{-\frac{q^2_\perp}{2 \kappa^2}} L_n^{|m|}(\frac{q_\perp^2}{\kappa^2}) e^{i m \theta_q}, \label{eq:transverse_basis}
  \\
  \chi_l(x; &\alpha, \beta) = x^{\frac{\beta}{2}}(1-x)^{\frac{\alpha}{2}}P_l^{(\alpha, \beta)}(2x-1)
  \nonumber\\
  \times&\sqrt{4\pi (2l+\alpha+\beta+1)}  \sqrt{\frac{\Gamma(l+1)\Gamma(l+\alpha+\beta+1)}{\Gamma(l+\alpha+1)(\Gamma(l+\beta+1)}}. \label{eq:longitudinal_basis}
\end{align}
In the transverse direction, we have the 2-dimensional harmonic oscillator function $\phi_{nm}(\textbf{q}_\perp)$, where $\textbf{q}_\perp \triangleq \mathbf{k}_\perp/\sqrt{x(1-x)}$, $q_\perp = |\textbf{q}_\perp|$, $\theta_q = \text{arg } \textbf{q}_\perp$, and $L_n^a(z)$ is the generalized Laguerre polynomial. The confining strength $\kappa$ serves as the harmonic oscillator scale parameter. Integers $n$ and $m$ are the principal quantum number for radial excitations and the orbital angular momentum projection quantum numbers, respectively. For the longitudinal basis function $\chi_l(x; \alpha, \beta)$, $l$ is the longitudinal quantum number and $P_l^{(\alpha, \beta)}(z)$ is the Jacobi polynomial. Parameters $\alpha$ and $\beta$ are dimensionless basis parameters in this model specified by $\alpha=2m_{\bar{q}}(m_q+m_{\bar{q}})/\kappa^2$ and $\beta=2m_q(m_q+m_{\bar{q}})/\kappa^2$.

To make the eigenvalue problem specified by the Hamiltonian in our basis representation numerically tractable, the basis modes are truncated to their respective finite cutoffs: $2n+|m|+1 \leq N_\mathrm{max}$ and $0 \leq l \leq L_\mathrm{max}$. $N_\mathrm{max}$ controls the total allowed oscillator quanta and it provides a natural infrared (IR) regulator $\Lambda_\mathrm{IR} \sim b/\sqrt{N_\mathrm{max}}$ and ultraviolet (UV) regulator $\Lambda_\mathrm{UV} \sim b \sqrt{N_\mathrm{max}}$~\cite{Li:2015zda}. $L_\mathrm{max}$ controls the resolution of the basis in the longitudinal direction.
 
%%%%%%%%%%%%%%% SECTION 3 %%%%%%%%%%%%%%%%
\section{Numerical Results}
In this work, we focus our investigation on the light-unflavored mesons without restricting the isospin and charge. We work in the $\mathrm{SU}(2)$ isospin symmetric limit such that the anti-quark mass $m_{\bar{q}}$ is identical to the quark mass $m_q$. We find the basis cutoff $N_\mathrm{max} = 8$ to be sufficient, which is identified with $\Lambda_\mathrm{IR}$ = 0.21 GeV and $\Lambda_\mathrm{UV}$ = 1.83 GeV. $L_\mathrm{max} = 24$ is used mainly to reduce the artifact of numerical oscillations in the longitudinal direction of the LFWFs. In the remainder of the paper, we will use $N_\mathrm{max} = 8$ and $L_\mathrm{max}=24$ for all calculations unless stated otherwise. We first apply the Hamiltonian $H_{\text{eff}}$ to find the optimal parameters of confining strength $\kappa$ and constitute quark mass $m_q$ by fitting the masses of $\rho$(770) and $\rho'$(1450) from the Particle Data Group (PDG)~\cite{PhysRevD.98.030001}. Next, we fine tune the coupling strength of pseudoscalar interaction $\lambda$ to best fit the experimental mass value for the $\pi$(140). This two-step procedure produces the optimized model parameters summarized in Table~\ref{tab:model_params}. 

\begin{table*}
  \centering
  \caption{Summary of the model parameters. The first line provides the best parameters when Eq. \eqref{eq:Heff} alone is used. The second line shows the modifications arising from the addition of the pseudoscalar interaction in Eq. \eqref{eq:Heff5}. The root mean square (r.m.s.) measures the overall deviations between the Particle Data Group (PDG)~\cite{PhysRevD.98.030001} values and BLFQ results for the 11 identified states shown in Fig.~\ref{fig:mass_spect}. The quantity $\overline{\delta_j M}$ is defined in the text.} 
  \begin{ruledtabular}
    \begin{tabular}{  l  c@{\hskip 0.1in}  c@{\hskip 0.15in}  c@{\hskip 0.1in}  c@{\hskip 0.1in}  c@{\hskip 0.1in}  c@{\hskip 0.15in}  c@{\hskip 0.1in}  c@{\hskip 0.1in}  c@{\hskip 0.1in}  c  }
    \\[-9pt]
    & $N_{\mathrm{f}}$ 
    & $\alpha_{\mathrm{s}}(0)$  
    & $\kappa \text{ (MeV)}$   
    & $m_q=m_{\bar{q}}$ (MeV)
    & $N_\mathrm{exp}$  
    & $N_\mathrm{max}$ 
    & $L_\mathrm{max}$ 
    & $\lambda \text{ (GeV}^{-2}\text{)}$ 
    & r.m.s. (MeV)
    & $\overline{\delta_j M}$ (MeV) 
    \\[6pt] \hline 
    \\[-6pt]
    $H_{\text{eff}}$           & \multirow{2}{*}{3} & \multirow{2}{*}{0.89} &  \multirow{2}{*}{610}  & \multirow{2}{*}{480} & \multirow{2}{*}{11}   & \multirow{2}{*}{8} & \multirow{2}{*}{24} & - & 127 & 61 \\[4pt]   
    $H_{\text{eff,$\gamma_5$}}$  &  &  &  &  &  &  &  & 0.56 & 111 & 60\\[5pt] 
    \end{tabular}
  \end{ruledtabular}
  \label{tab:model_params}
\end{table*}

\subsection{Spectroscopy}
\begin{figure*}
  \centering
  \subfigure{\includegraphics[width=0.48\linewidth]{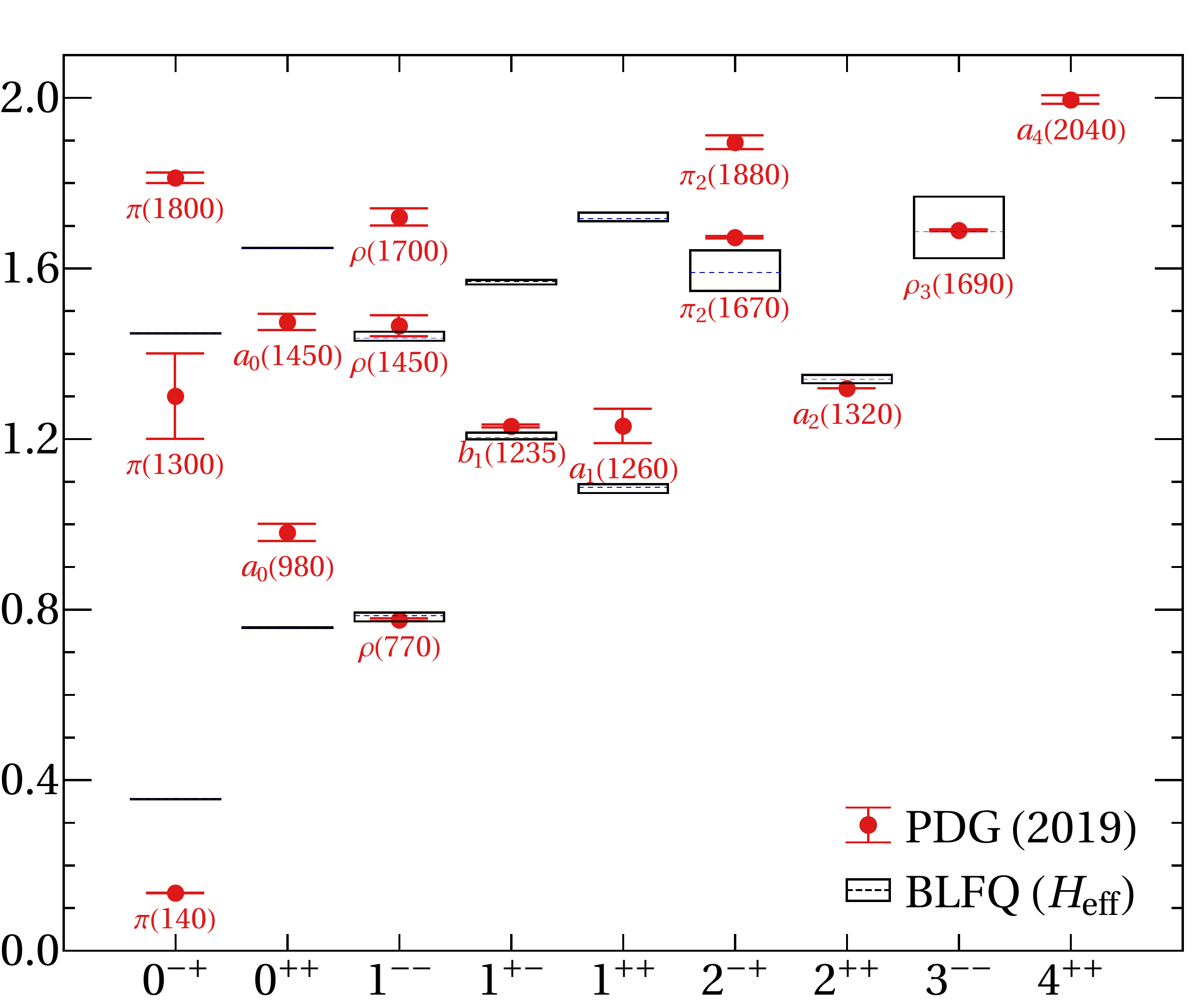}}\quad\quad
  \subfigure{\includegraphics[width=0.48\linewidth]{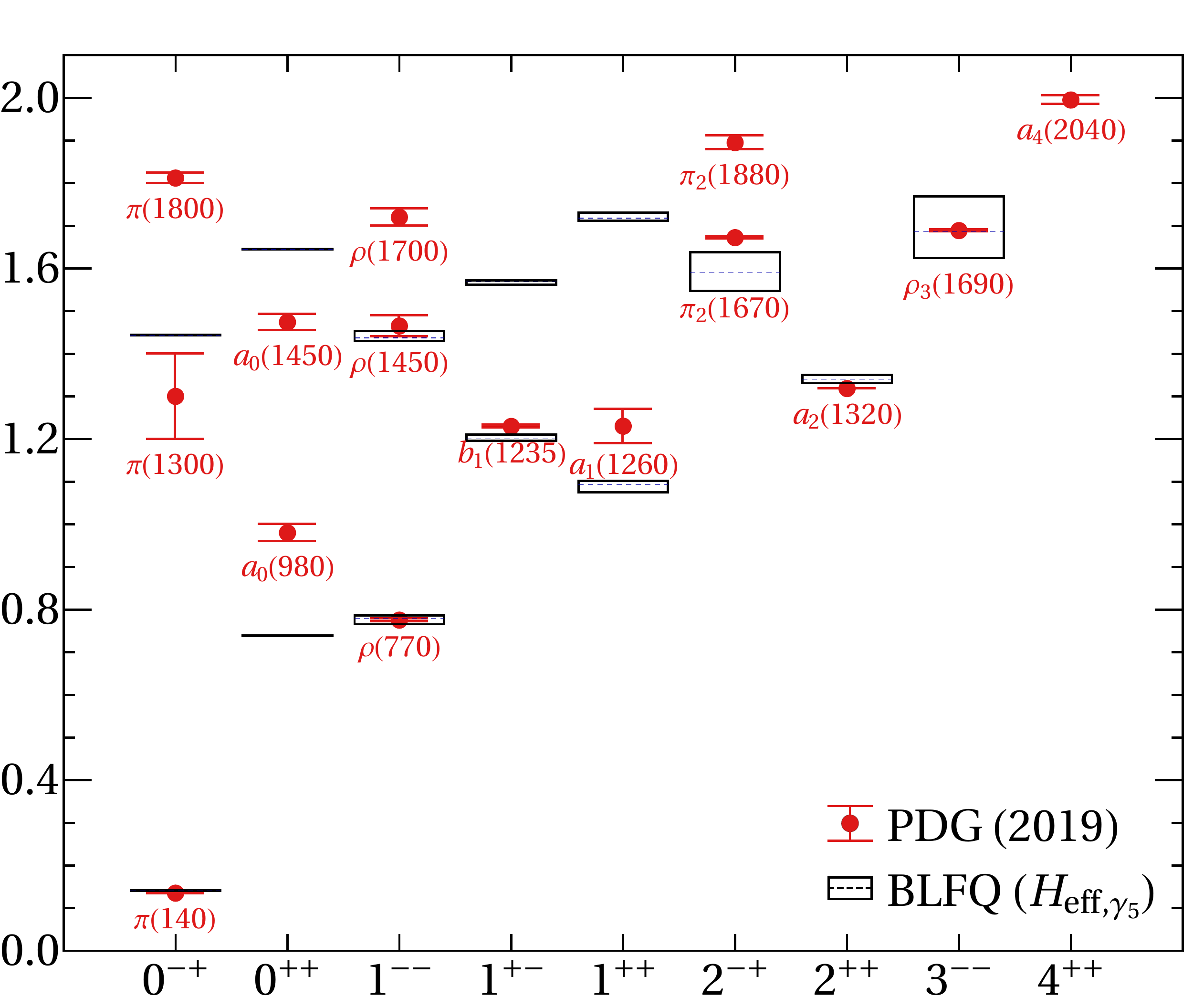}}
  \caption{\label{fig:mass_spect}The reconstructed light meson spectra with effective Hamiltonians $H_\text{eff}$ (\textit{left panel}) and $H_\text{eff, $\gamma_5$}$ (\textit{right panel}) at $N_\mathrm{max}=8$ and $L_\mathrm{max}=24$. The horizontal and vertical axes are $j^\text{PC}$ and invariant mass in GeV respectively. Model parameters are listed in Table \ref{tab:model_params}. Calculated states are represented by boxes to show the spread of mass eigenvalues in $m_j$ due to violation of rotational symmetry and basis truncations. The r.m.s. deviations of the masses from the PDG values are 127 MeV (left panel) and 111 MeV (right panel) for the 11 identified states both illustrated in this figure and specified in Table~\ref{tab:spectroscopy}. We use BLFQ ($H_\text{eff, $\gamma_5$}$) to indicate the model with both the one-gluon exchange interaction and the pseudoscalar interaction, and BLFQ ($H_\mathrm{eff}$) to indicate the model without the pseudoscalar interaction.}
\end{figure*}

The effective Hamiltonian in Eq.~\eqref{eq:eigenvalue_equation} is diagonalized separately for each value of $m_j$. We then perform state identification to assign the quantum numbers of the states, $j^{\text{PC}}$, where $j$ is the total angular momentum, P is the parity, and C is the charge conjugation, as presented in Fig.~\ref{fig:mass_spect}. Light-front parity (or $(-1)^j$P) as well as C can be deduced from the basis representation of the LFWFs~\cite{Li:2015zda}, while $j$ is determined from the mass degeneracy of the spectrum. In these figures, we use the tops and bottoms of the black boxes to indicate the spreads of mass eigenvalues of the same states with different $m_j$. The mean values marked by the dashed bars are defined as
\begin{align}
 \overline{M} \triangleq \sqrt{\frac{M_{-j}^2+ M_{1-j}^2+\cdots + M_j^2}{2j+1}}.
\end{align}
In addition, we use $\delta_j M_h \triangleq \text{max}\, \{M_{m_j}\} - \text{min}\,\{M_{m_j}\}$ for each hadron $h$ to measure its violation of rotational symmetry, represented by the height of the boxes. The overall violation of rotational symmetry for the entire spectrum is defined as
\begin{align}
  \overline{\delta_j M} \triangleq \sqrt{\frac{1}{N}\sum_h (\delta_j M_h)^2},
\end{align}
where $N$ is number of hadrons with $j\ne 0$ and the summation is over all such hadrons identified in the model.

\begin{table}
  \centering
  \caption{Detailed comparison of the mass measured in GeV between the PDG~\cite{PhysRevD.98.030001} and BLFQ results for the 11 identified states illustrated in Fig.~\ref{fig:mass_spect}. Results of AdS/QCD~\cite{holography} use a quark mass around 46 MeV and confining strength around 540 MeV. Results of BSE calculations with the rainbow-ladder truncation~\cite{Fischer:2014xha} are also listed, along with the improved results using the three-loop truncation of the three-particle irreducible effective action shown in the parenthesis~\cite{Williams:2015cvx}. Mean values are used for both experimental and theoretical data. } 
  \begin{ruledtabular}
    \begin{tabular}{r@{\hskip 0.25in} c@{\hskip 0.18in}  c@{\hskip 0.2in}  c@{\hskip 0.24in} c@{\hskip 0.29in}}
    \\[-10pt]
    & PDG~\cite{PhysRevD.98.030001}
    & AdS/QCD~\cite{holography}
    & BSE~\cite{Fischer:2014xha, Williams:2015cvx} 
    & BLFQ
    \\[6pt] \hline 
    \\[-5pt]
    $\pi$        &  0.14    & 0.14     & 0.14 (0.14)   & 0.14     \\[4pt]   
    $\rho$       &  0.78    & 0.78     & 0.76 (0.74)   & 0.78     \\[4pt]   
    $a_0$        &  0.98    & 0.78     & 0.64 (1.1)    & 0.74     \\[4pt]   
    $b_1$        &  1.23    & 1.09     & 0.85 (1.3)    & 1.20    \\[4pt]   
    $a_1$        &  1.23    & 1.09     & 0.97 (1.3)    & 1.09    \\[4pt]   
    $\pi'$       &  1.30    & 1.09     & 1.10          & 1.44    \\[4pt]   
    $a_2$        &  1.32    & 1.33     & 1.16          & 1.34    \\[4pt]   
    $\rho'$      &  1.45    & 1.33     & 1.02          & 1.44    \\[4pt]   
    $a_0'$       &  1.47    & 1.33     & 1.27          & 1.65    \\[4pt]   
    $\pi_2$      &  1.67    & 1.53     & 1.23          & 1.59    \\[4pt]   
    $\rho_3$     &  1.69    & 1.71     & 1.54          & 1.69    \\[5pt]   
    \end{tabular}
  \end{ruledtabular}
  \label{tab:spectroscopy}
\end{table}

We can see that the effective Hamiltonian $H_\text{eff, $\gamma_5$}$ is successful in reproducing the low-lying experimental masses in the light sector. The spectroscopy is mostly rotational symmetric with $\overline{\delta_j M}$ around 60 MeV, as shown in Table~\ref{tab:model_params}. Moreover, the addition of the pseudoscalar interaction generates sufficient mass splitting between the pseudoscalar and the vector ground states, which serves to simulate the chiral dynamics sufficiently while holding the mass of other states primarily unchanged from the results with $H_\mathrm{eff}$ alone. For the rest of this work, we focus on the numerical results obtained using LFWFs from the BLFQ ($H_{\mathrm{eff}, \gamma_5}$) Hamiltonian unless stated otherwise. 

\subsection{Light-Front Wave Functions}
The valence structure of bound states is encoded in the LFWFs, through which various hadronic observables are directly accessible. To visualize the LFWF as the probability amplitude, we use $k_\perp$ to sample $\mathbf{k}_{\perp}$ along the $x$-direction (i.e. $k_\perp = \mathbf{k}_{\perp} \cdot \hat{\mathbf{x}}$) and obtain a relative sign of $\exp(i m_l \pi) = (-1)^{m_l}$ for negative $k_{\perp}$. We also define $\psi^{(m_j)}_{\uparrow\downarrow \pm \downarrow\uparrow}(k_\perp, x) \triangleq \big [\psi^{(m_j)}_{\uparrow\downarrow}(k_\perp, x) \pm \psi^{(m_j)}_{\downarrow\uparrow}(k_\perp, x) \big] / \sqrt{2}$. For pseudoscalars ($0^-$ states), there are two independent components: $\psi_{\uparrow\downarrow - \downarrow\uparrow}(k_\perp, x) $ and $\psi_{\uparrow\uparrow}(k_\perp, x)  =  \psi_{\downarrow\downarrow}(k_\perp, x)$. For vectors ($1^-$ states) at $m_j=0$, there are also two independent components: $\psi_{\uparrow\downarrow + \downarrow\uparrow}(k_\perp, x) $ and $\psi_{\uparrow\uparrow}(k_\perp, x)  =  \psi_{\downarrow\downarrow}(k_\perp, x)$.

\begin{figure*}
  \centering
  \subfigure[\ $\eta_b$: $\psi_{\uparrow\downarrow - \downarrow\uparrow}(k_\perp, x)$]
  {\includegraphics[width=0.32\linewidth]{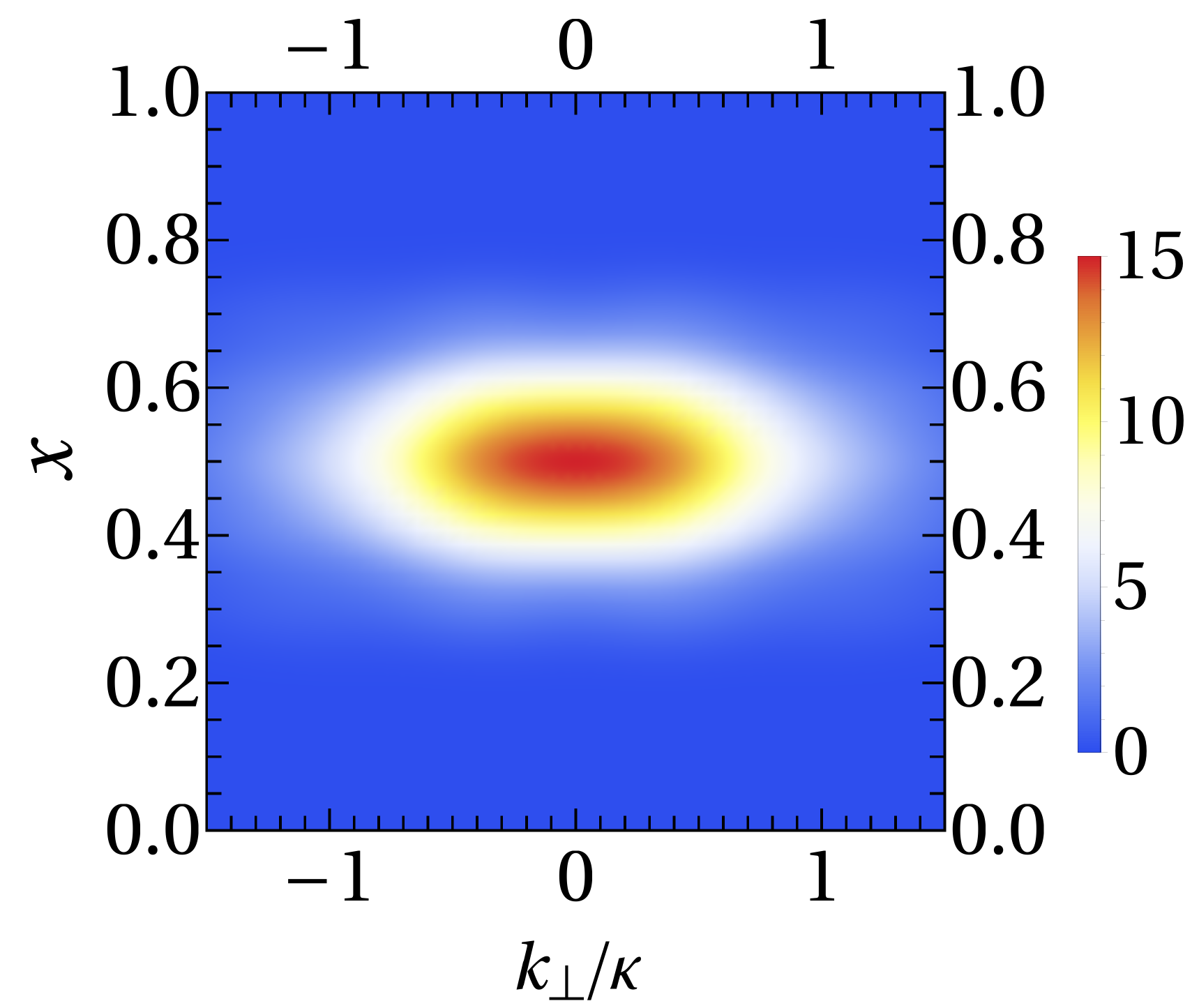}}\quad
  \subfigure[\ $\eta_c$: $\psi_{\uparrow\downarrow - \downarrow\uparrow}(k_\perp, x)$]
  {\includegraphics[width=0.32\linewidth]{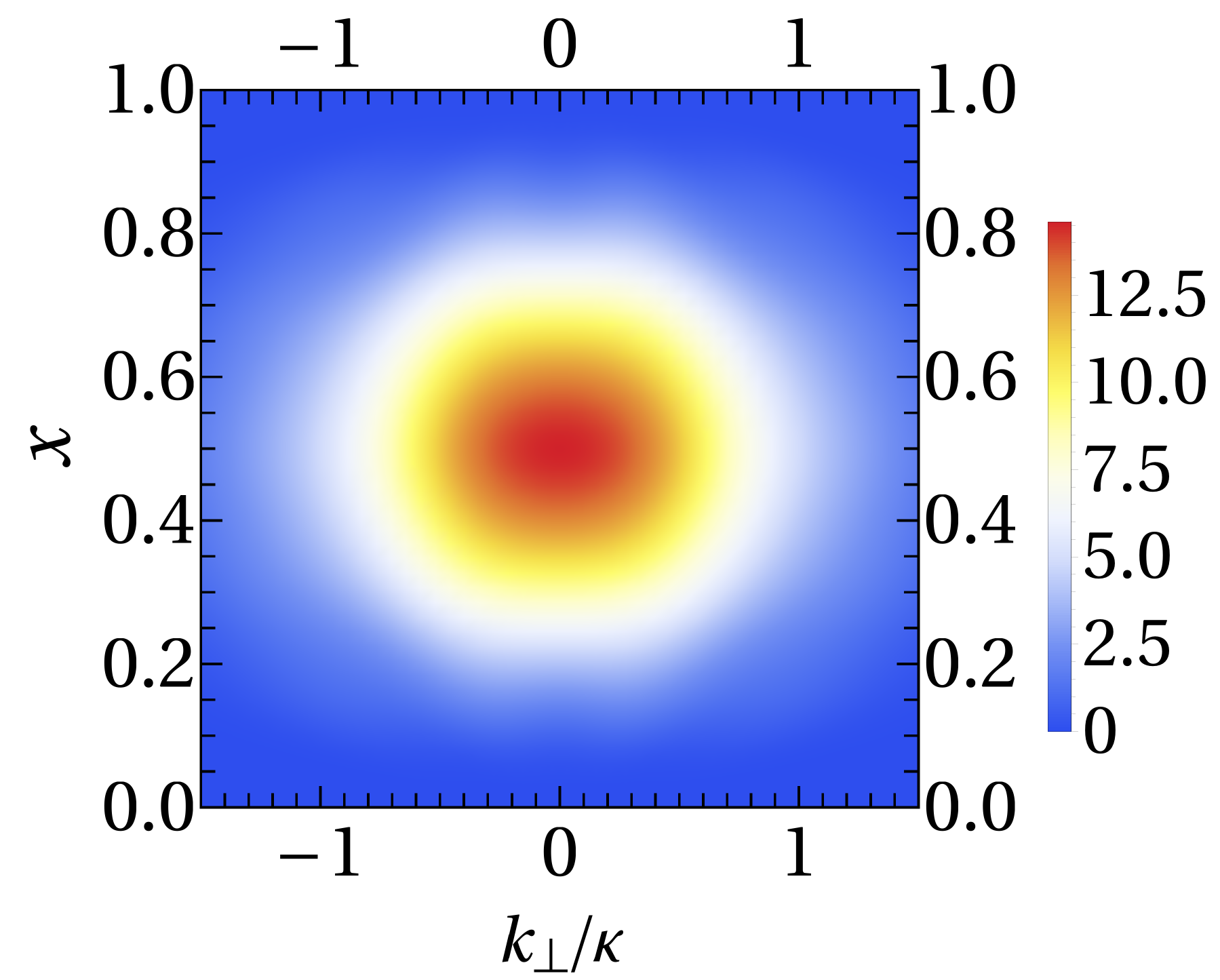}}\quad
  \subfigure[\ $\pi$: $\psi_{\uparrow\downarrow - \downarrow\uparrow}(k_\perp, x)$]
  {\includegraphics[width=0.32\linewidth]{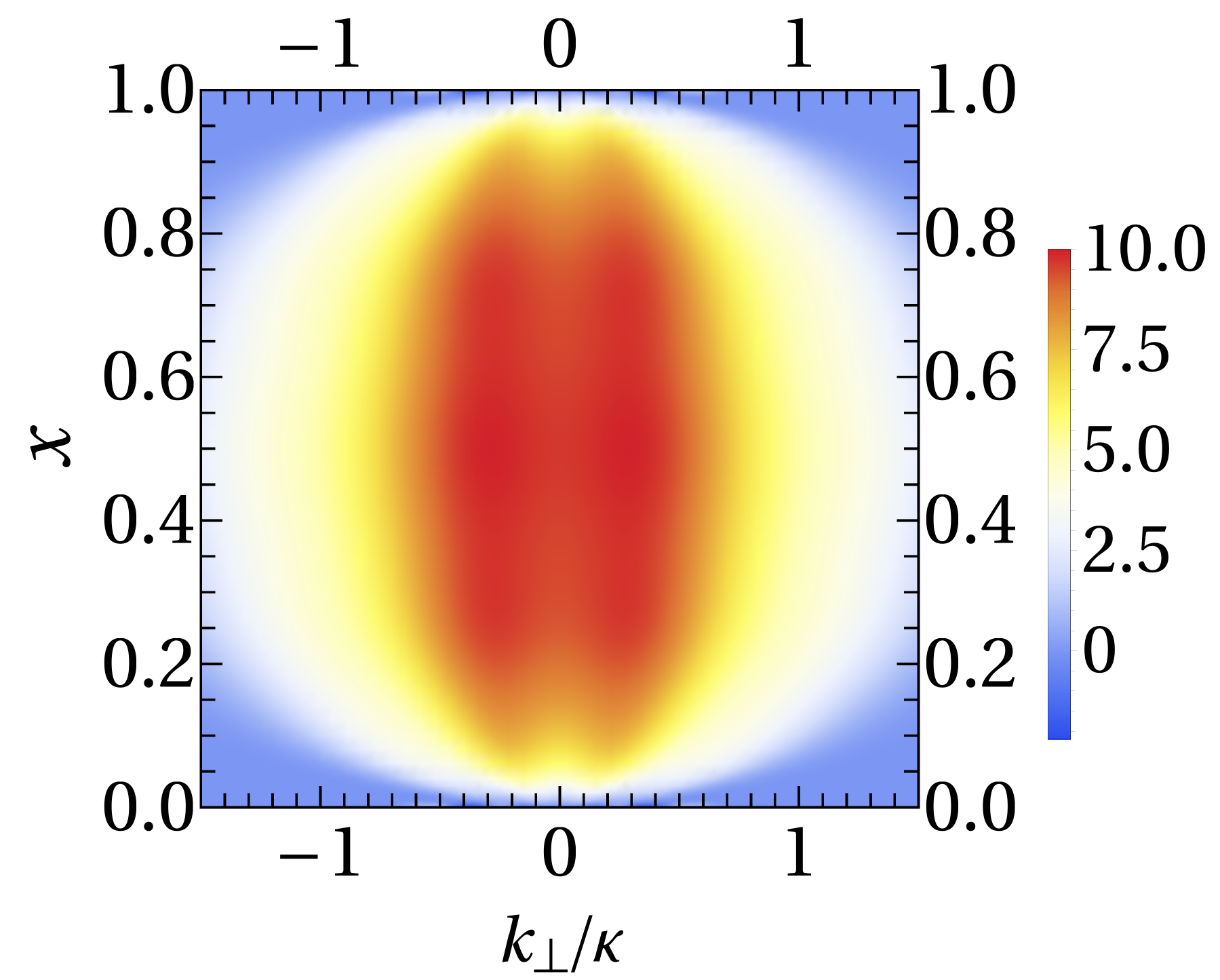}}
  \caption{\label{fig:evo_ps}Evolution of dominant LFWF $\psi_{\uparrow\downarrow - \downarrow\uparrow}(k_\perp, x) $ for pseudoscalar mesons $\eta_b$, $\eta_c$ and $\pi$ respectively (from left to right) at $N_\mathrm{max}=L_\mathrm{max} = 8$ with BLFQ ($H_\text{eff}$) Hamiltonian using different model parameters. The corresponding confining strengths $\kappa$ (quark mass $m_q$) are, from left to right, 1.387 GeV (4.894 GeV), 0.985 GeV (1.570 GeV) and 0.610 GeV (0.480 GeV). See Ref.~\cite{Li:2017mlw} for a complete list of parameters used for heavy quarkonia.}
\end{figure*}

It is instructive to compare LFWFs for the light mesons to those for the heavy quarkonia~\cite{Li:2017mlw} all obtained from the same Hamiltonian ($H_\mathrm{eff}$ of Eq. \eqref{eq:Heff}) but with different parameters. In Fig.~\ref{fig:evo_ps}, we show the comparison of the dominant pseudoscalar LFWF component ($\uparrow\downarrow - \downarrow\uparrow$) among $\eta_b$ (bottomonium), $\eta_c$ (charmonium), and $\pi$. Similar observations, though not shown here, can also be made about the evolution of the dominant vector LFWF component ($\uparrow\downarrow + \downarrow\uparrow$) in the $m_j=0$ sector among the $\Upsilon$, $J$/$\Psi$, and $\rho$ mesons. In these plots, the horizontal axes are the transverse momentum scaled by its respective confining strength which is the same as the basis scale used in the models. We notice that, with decreasing meson mass, the LFWFs span a broader kinematic region in both $x$ and $k_\perp/\kappa$ for the light mesons, as expected due to the anticipated increase in relativistic behavior.

\begin{figure*}
  \centering
  \subfigure[\ $\pi$: $\psi_{\uparrow\downarrow - \downarrow\uparrow}(k_\perp, x)$]{
    \includegraphics[width=0.32\linewidth]{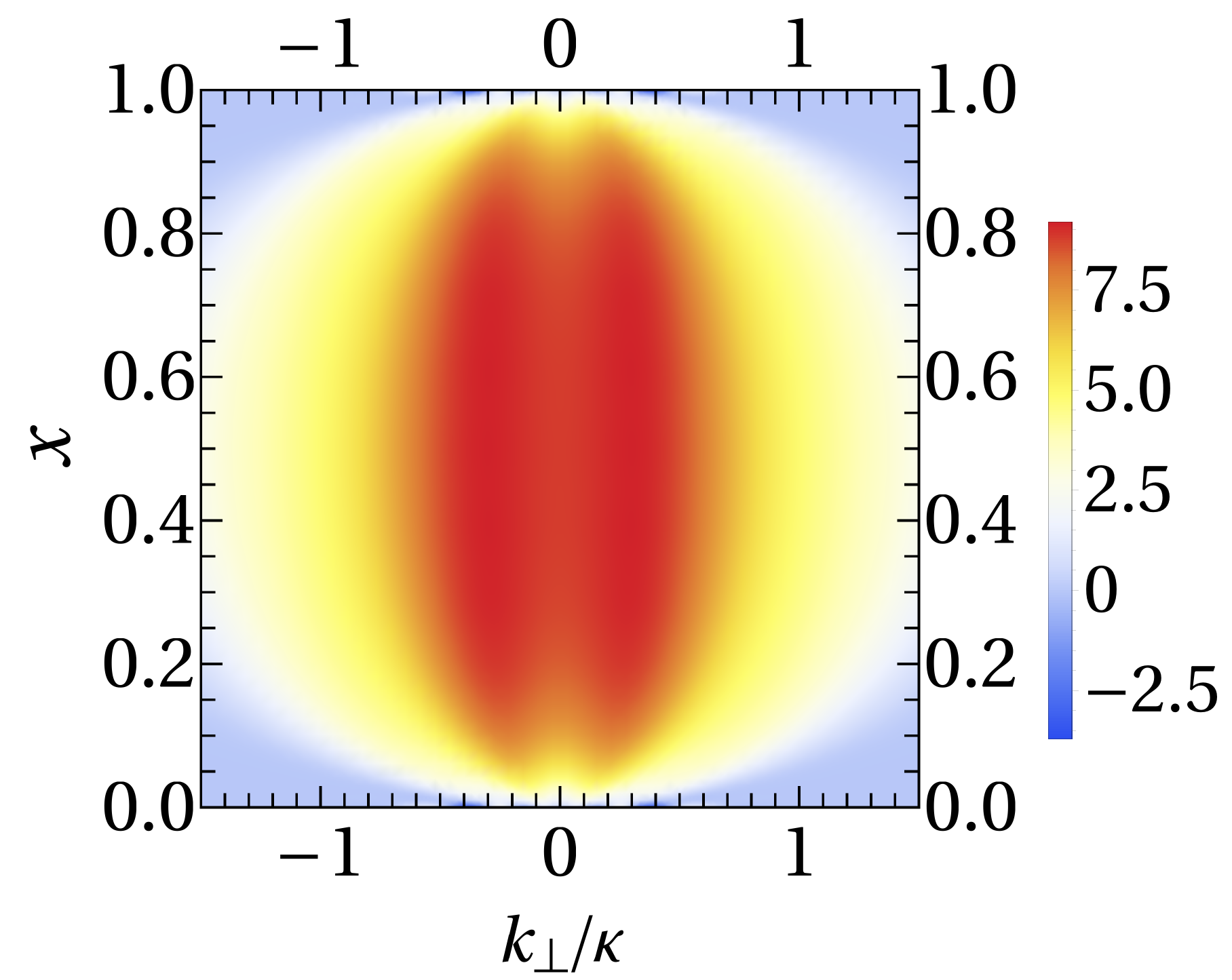}}\quad\quad
  \subfigure[\ $\pi$: $\psi_{\uparrow\uparrow}(k_\perp, x)$ = $\psi_{\downarrow\downarrow}(k_\perp, x)$]{
    \includegraphics[width=0.32\linewidth]{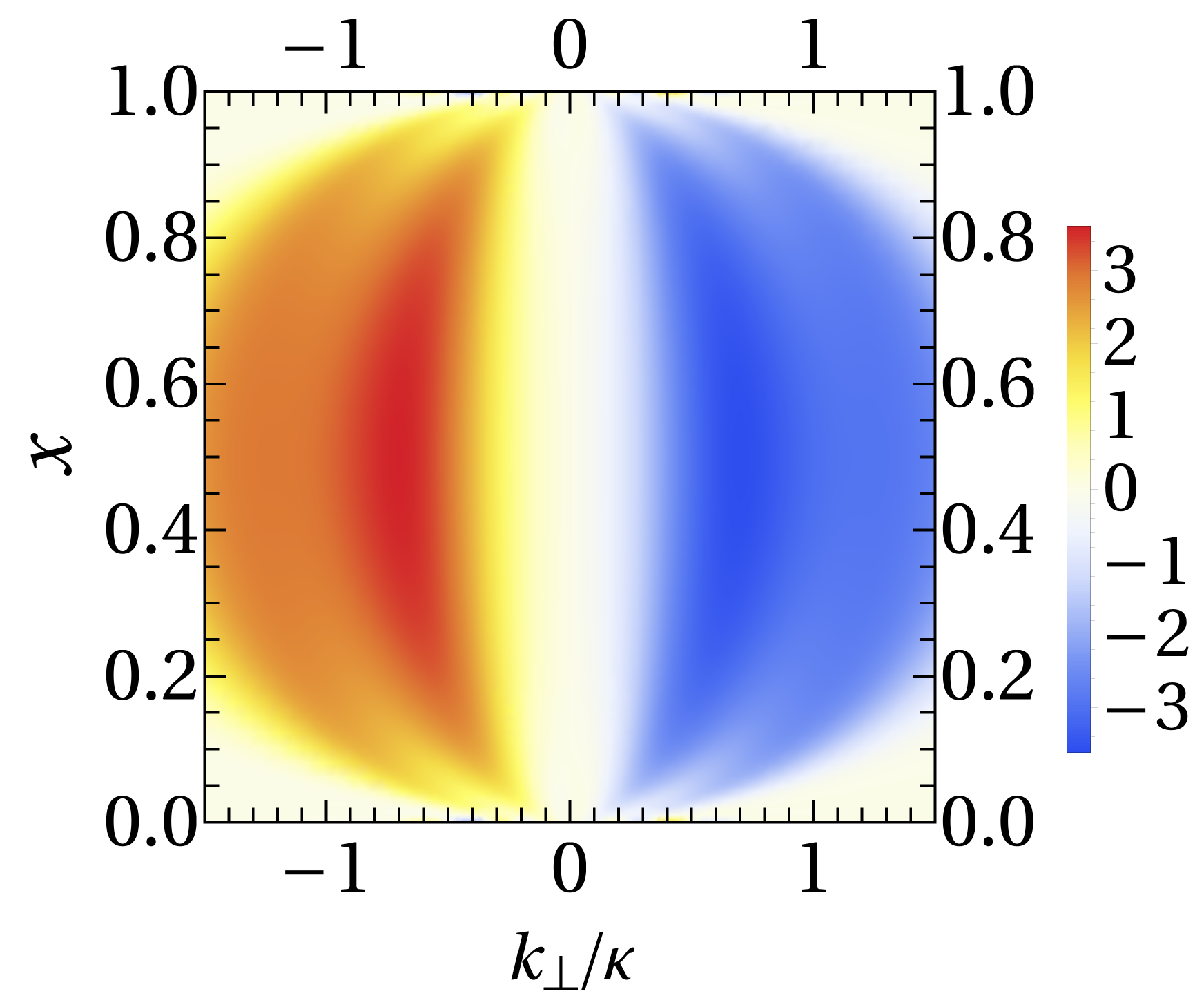}}
  \subfigure[\ $\rho$: $\psi_{\uparrow\downarrow + \downarrow\uparrow}(k_\perp, x)$]{
    \includegraphics[width=0.32\linewidth]{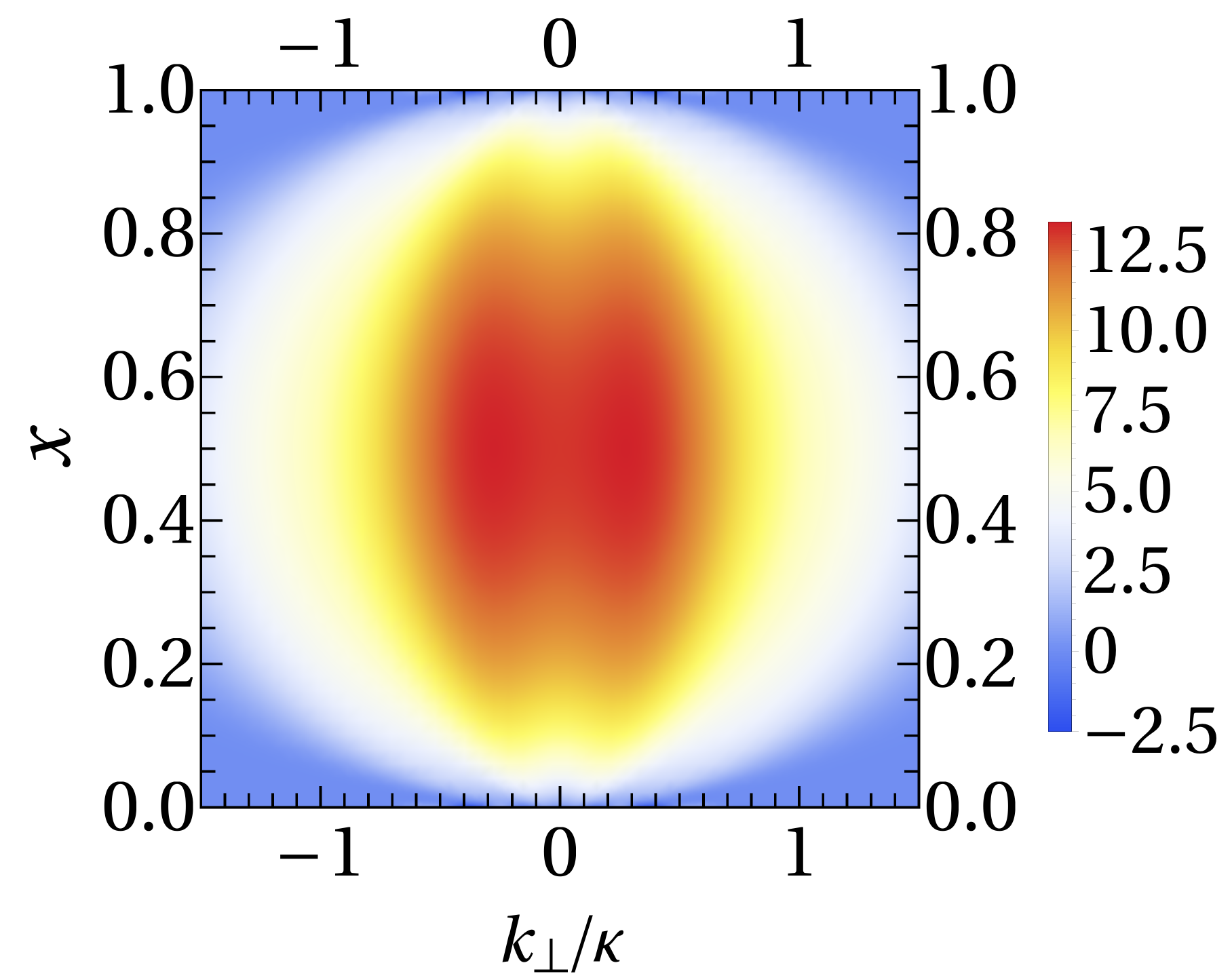}}\quad\quad
  \subfigure[\ $\rho$: $\psi_{\uparrow\uparrow}(k_\perp, x)$ = $\psi_{\downarrow\downarrow}(k_\perp, x)$]{
    \includegraphics[width=0.32\linewidth]{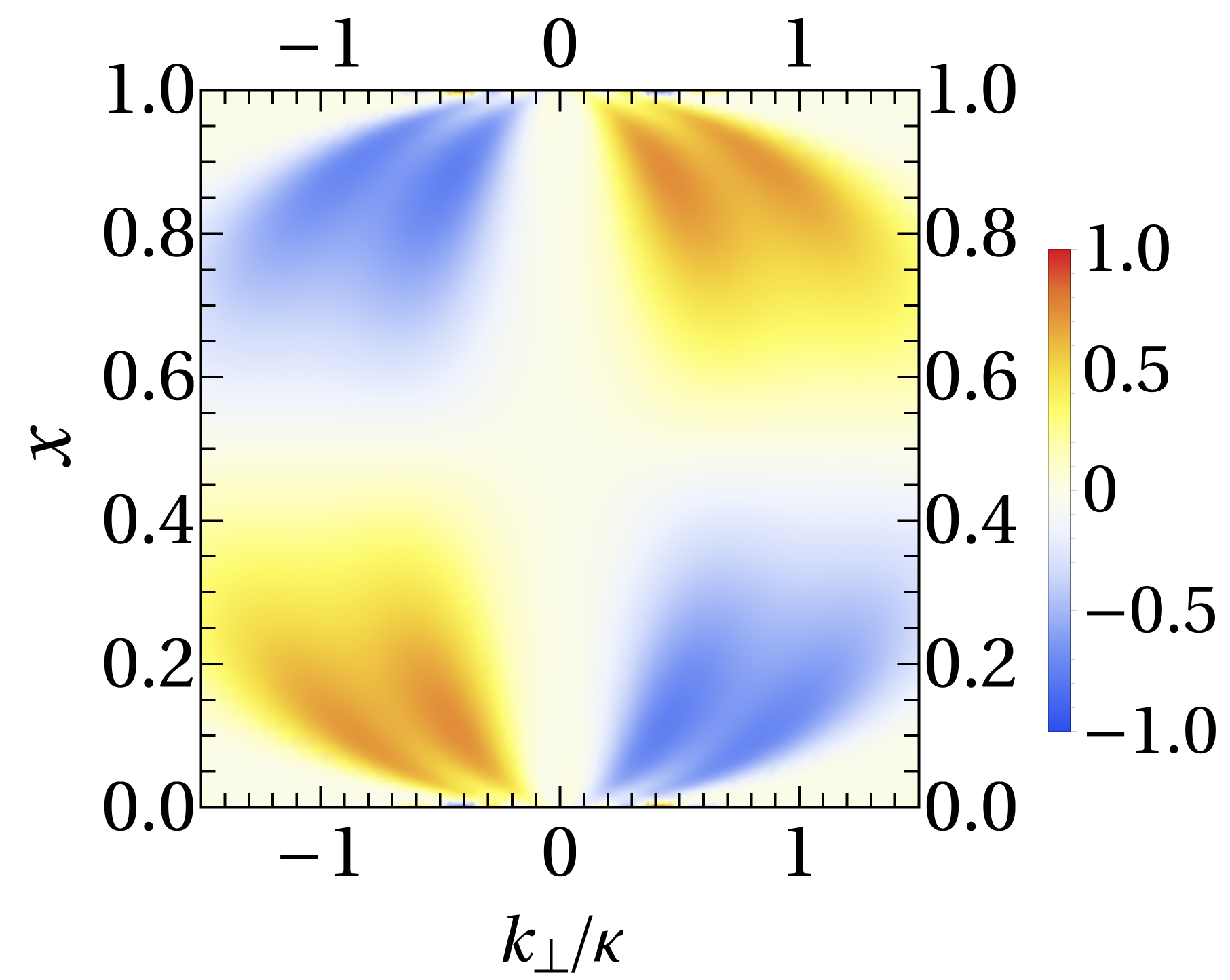}}
  \caption{\label{fig:LFWF_psrun}LFWF spin components of the pion and the rho meson ($m_j=0$) calculated using the Hamiltonian BLFQ ($H_\text{eff, $\gamma_5$}$) at $N_\mathrm{max}=8$ and $L_\mathrm{max} = 24$.}
\end{figure*}

With the addition of the pseudoscalar interaction, in Fig.~\ref{fig:LFWF_psrun}, we show the nonvanishing LFWF spin components of $\pi$ and $\rho$ at $m_j=0$ using $N_\mathrm{max}=8$ and $L_\mathrm{max} = 24$. It should be mentioned that the dominant pseudoscalar LFWF component contributes to around 62.9\% of the total amplitude in $\pi$, in constrast to 92.6\% in $\eta_c$ and 97.6\% in $\eta_b$, which further indicates the relativistic nature of the light meson system.

\subsection{Decay Constants}
Decay constants are defined from the local vacuum-to-hadron matrix elements of the quark current operators. For pseudoscalar (P) and vector (V) meson states, the decay constants $f_\mathrm{P}$ and $f_\mathrm{V}$ are defined as
\begin{align}
\langle 0 | \bar{\psi}(0) \gamma^\mu \gamma_5 \psi(0)|\mathrm{P}(p)\rangle &= i p^\mu f_{\mathrm{P}}, \label{eq:decay1}\\
\langle 0 | \bar{\psi}(0) \gamma^\mu  \psi(0)|\mathrm{V}(p, m_j)\rangle &= e_{m_j}^{\mu}(p) M_{\mathrm{V}} f_{\mathrm{V}} \label{eq:decay2},
\end{align}
where $p$ is the meson momentum, $M_\mathrm{V}$ is the mass of the vector meson, and $e_{m_j}^{\mu}(p)$ is the polarization vector (we adopt the convention of Ref.~\cite{Li:2015zda}). We take the $m_j=0$ state of the vector meson for calculating the decay constant. The ``+" and transverse ($x, y$) currents lead to the same result in the leading Fock sector~\cite{ML_thesis}. 
\begin{align}\label{eq:decay}
\frac{f_\text{P,V}}{2\sqrt{2N_c}} = \int^1_0 \frac{dx}{2\sqrt{x(1-x)}} \int \frac{d^2 \mathbf{k}_\perp}{(2\pi)^3} \psi^{(m_j = 0)}_{\uparrow\downarrow \mp \downarrow\uparrow}(x, \mathbf{k}_\perp).
\end{align}
In Table~\ref{tab:decay_const}, we present the results of the decay constants for $\pi$ and $\rho$ to compare with experiments and other methods. Our results of decay constants are consistently larger than the PDG data~\cite{PhysRevD.98.030001}, which is primarily due to the fact that the model treats the valence quarks as point-like particles. In principle, this can be improved by incorporating higher Fock sectors which, among other effects, will bring in the self-energy contributions to the quarks and anti-quarks. In addition, we find the decay constant of the excited pion to be $f_\pi' = 136$ MeV and that of the excited rho meson to be $f_\rho' = 133$ MeV in our work. 
\begin{table}
\centering
\caption{Decay constants (in MeV) of ground states $\pi$ and $\rho$ calculated according to Eq.~\eqref{eq:decay} with BLFQ ($H_\text{eff, $\gamma_5$}$) at $N_\mathrm{max}=8$ and $L_\mathrm{max}=24$ compared with experiments and other theoretical models. For the BLFQ-NJL model, we present both the original results from Ref.~\cite{Jia} and the revised calculations in parentheses~\cite{Jia_revised}. For the light-front quark model (LFQM), we show results of both harmonic oscillator (left) and linear (right) confining potentials. 
} 
\begin{ruledtabular}
  \begin{tabular}{ c@{\hskip 0.05in}  c@{\hskip 0.05in}  c@{\hskip 0.05in}  c@{\hskip 0.05in}  c@{\hskip 0.05in}  c@{\hskip 0.15in}}
  \\[-10pt]
  & PDG~\cite{PhysRevD.98.030001} 
  & This work
  & BLFQ-NJL~\cite{Jia} 
  & BSE~\cite{Maris} 
  & LFQM~\cite{Choi:pion_rho} 
  \\[6pt] \hline 
  \\[-6pt]
                      % PDG    % BLFQ   % BLFQ-NJL  % BSE    % LFQM
  $f_\pi$           & 130    & 259    & ~202.10 (142.91)    & 131    & 130, 131      \\[4pt]
  $f_\rho$          & 216    & 323    & 100.12 (70.80)      & 207    & 246, 215      \\[6pt]
  \end{tabular}
\end{ruledtabular}
\label{tab:decay_const}
\end{table}
 
\subsection{Form Factors and Charge Radii}
The electromagnetic form factors characterize the structure of a bound state system in quantum field theory, which generalize the multipole expansion of the charge and current densities in nonrelativistic quantum mechanics. Here we study the fictitious form factor with the photon coupling only to the quark (but not the anti-quark). The form factors can be obtained by applying the Drell-Yan-West formula within the Drell-Yan frame $P'^+=P^+$~\cite{DrellYan}, which in the valence Fock sector depend on the matrix elements
\begin{align}
 I_{m_j', m_j} (Q^2) &\triangleq \langle \psi_h(P',j,m_j')  | \frac{J^+(0)}{2P^+}| \psi_h(P,j,m_j) \rangle 
 \nonumber \\
 &= \sum_{s, \bar{s}} \int_0^1 \frac{dx}{2x(1-x)}\int \frac{d^2\text{k}_\perp}{(2\pi)^3} 
 \nonumber\\
 &\quad \times\psi^{(m_j')*}_{s\bar{s}/h}(\mathbf{k}_\perp+(1-x)\textbf{q}_\perp, x)  \psi^{(m_j)}_{s\bar{s}/h}(\mathbf{k}_\perp, x) ,\label{eq:ff}
\end{align}
with $q = P'- P$ and $Q^2 = -q^2 = \textbf{q}^2_\perp$. For pseudoscalar mesons, Eq.~\eqref{eq:ff} directly produces the charge form factor $G_{\mathrm{C}}(Q^2) = I_{0,0}(Q^2)$. For vector mesons, we adopt the prescription of Grach and Kondratyuk~\cite{GK_1984, GK_1995} to specify the angular condition. In addition, wave functions of the same vector meson solved from the Hamiltonian might have relative sign differences across different $m_j$, so we have used the nonrelativistic dominant component to fix their relative signs in the calculation of form factor matrix elements. Consequently, the elastic form factors of the vector states, namely the charge form factor $G_{\mathrm{C}}(Q^2)$, the magnetic form factor $G_{\mathrm{M}}(Q^2)$, and the quadruple form factor $G_{\mathrm{Q}}(Q^2)$ are given by
\begin{equation}
  \begin{aligned}\label{eq:ff_vector}
  G_{\mathrm{C}}(Q^2) &= \frac{1}{3} \big [(3-2\eta) \ I_{1, 1} + 2\sqrt{2\eta} \ I_{1,0} + I_{1,-1} \big ],  \\
  G_{\mathrm{M}}(Q^2) &= 2 \ I_{1,1} - \sqrt{\frac{2}{\eta}} \ I_{1,0}, \\
  G_{\mathrm{Q}}(Q^2) &= -\frac{2}{3} \sqrt{2} \big [ \ \eta \ I_{1, 1} - \sqrt{2\eta} \ I_{1,0} + I_{1,-1} \big ],  
  \end{aligned}
\end{equation}
with $\eta = Q^2/ (4M_h^2)$ and $M_h$ is the mass of the hadron. We then define the charge r.m.s. radius $\sqrt{\langle r^2 \rangle}$, the magnetic moment $\mu$, and the quadrupole moment $\mathcal{Q}$ as follows:
\begin{equation}
  \begin{aligned}\label{eq:ff_obs}
  \langle r^2 \rangle &=  -6  \lim_{Q^2 \to 0}\frac{\partial G_{\mathrm{C}}(Q^2)}{\partial Q^2} , \\[5pt]
  \mu &= \lim_{Q^2 \to 0}G_{\mathrm{M}}(Q^2), \\
  \mathcal{Q} &= 3 \sqrt{2} \lim_{Q^2 \to 0}\frac{\partial G_{\mathrm{Q}}(Q^2) }{\partial Q^2}.
  \end{aligned}
\end{equation}

\begin{figure*}
  \centering
  \subfigure{\includegraphics[width=0.48\linewidth]{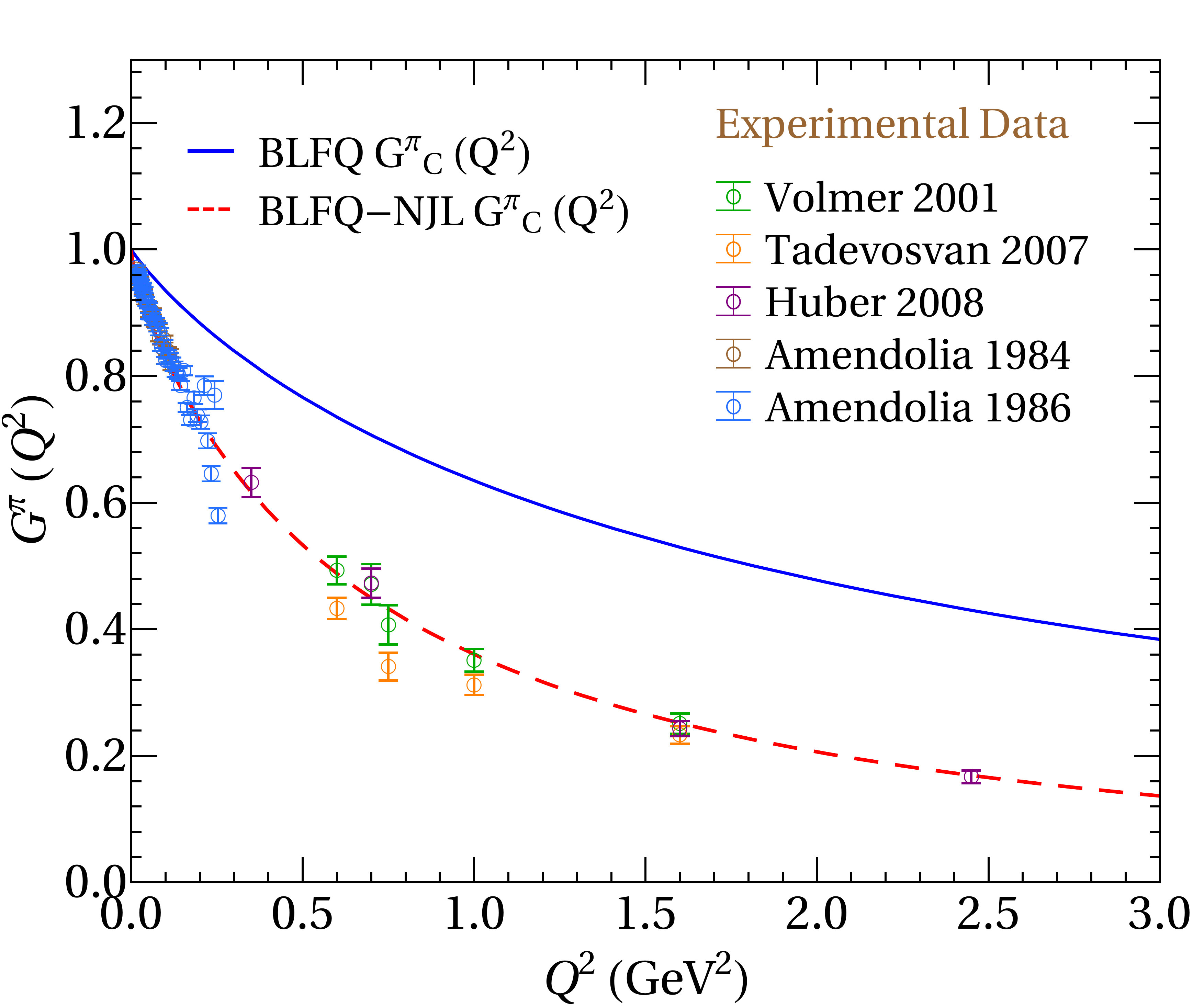}}\quad\quad
  \subfigure{\includegraphics[width=0.48\linewidth]{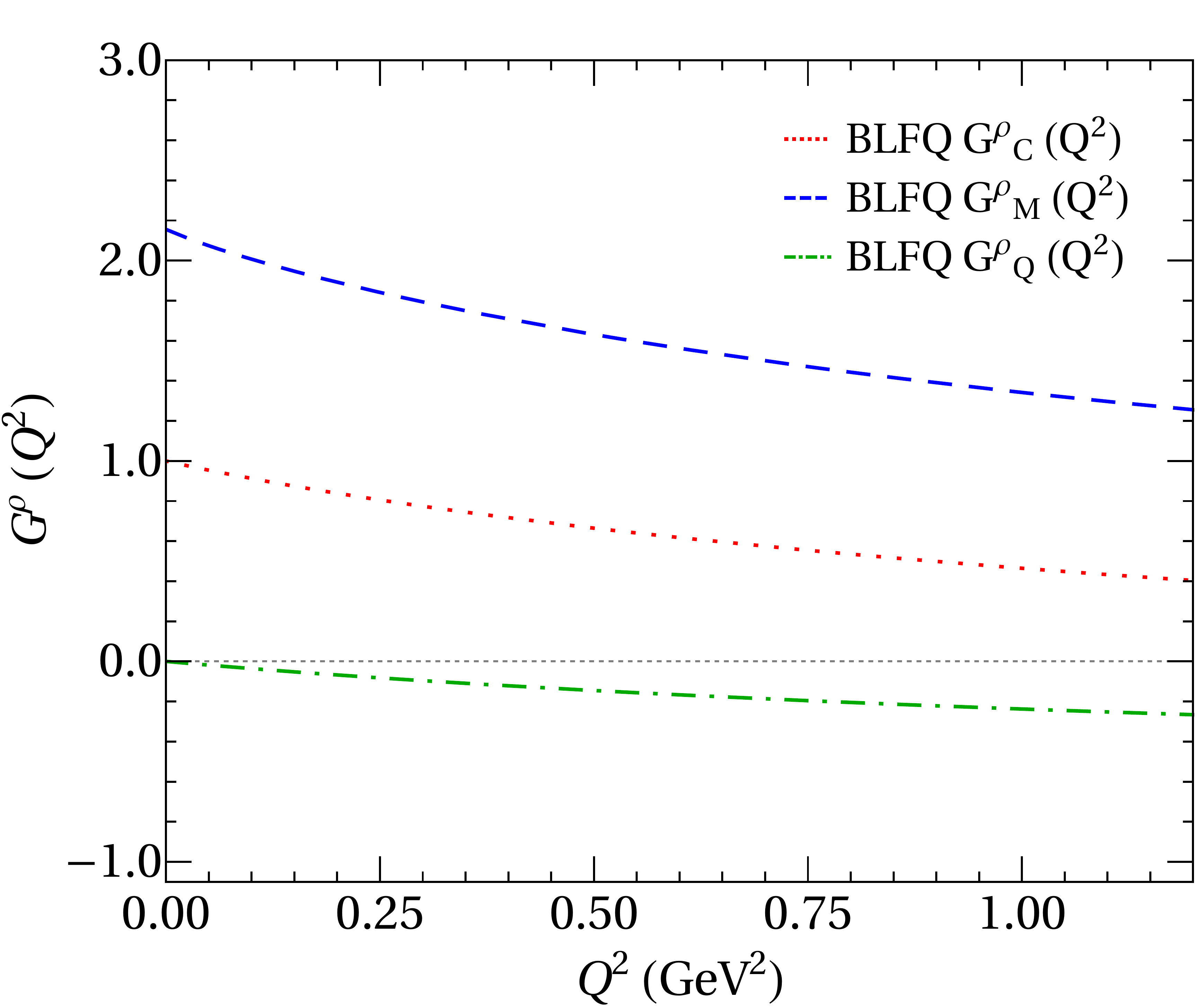}}
\caption{Elastic form factors for $\pi$ (\textit{left panel}) and $\rho$ meson (\textit{right panel}) as functions of $Q^2$ calculated according to Eqs.~\eqref{eq:ff} and~\eqref{eq:ff_vector} using LFWFs obtained from BLFQ ($H_\text{eff, $\gamma_5$}$) at $N_\mathrm{max}=8$ and $L_\mathrm{max}=24$. Experimental measurements of pion charge form factors are available~\cite{Volmer, Amendolia_1984, Amendolia_1986} and plotted on the left for comparison. The BLFQ-NJL model calculation of the charge form factor for the pion (where the charge radius was fit to experiment) is also plotted in the dashed line for comparison~\cite{Jia}. The charge form factor, magnetic form factor, and the quadruple form factors calculated from the rho meson LFWFs are illustrated on the right in the dotted line, the dashed line and the dot-dashed line, respectively.}
\label{fig:form_factors}
\end{figure*}

In Fig.~\ref{fig:form_factors}, we present the calculations of elastic form factors for $\pi$ and $\rho$ to compare with available experimental results as well as BLFQ-NJL results from Ref.~\cite{Jia}. Results of the observables defined in Eq. \eqref{eq:ff_obs} are included in Table~\ref{tab:ff_obs}. We can see our charge radii for $\pi$ and $\rho$ are lower than those from other approaches and experiments whenever available, which is consistent with the large decay constants found in our model (see Table~\ref{tab:decay_const}).

\begin{table*}
\centering
\caption{Results of the $\pi$ and $\rho$ meson charge radii $\sqrt{\langle r^2_h \rangle}$ (in $\text{fm}$), magnetic moments $\mu$ and quadrupole moments $\mathcal{Q}$ (in $\text{fm}^2$) calculated according to Eq.~\eqref{eq:ff_obs}, compared to available experimental measurements and other models. The lightest pion mass used in the lattice calculation in Ref.~\cite{Owen:2015} is $m_\pi = 161$ $\text{MeV}$. Uncertainties quoted in the referenced works are quoted in parenthesis. } 
\begin{ruledtabular}
  \begin{tabular}{ c@{\hskip 0.2in}  c@{\hskip 0.2in}  c@{\hskip 0.2in}  c@{\hskip 0.2in}  c@{\hskip 0.2in}  c@{\hskip 0.2in}  c@{\hskip 0.2in} c  }
  \\[-9pt]
  & PDG~\cite{PhysRevD.98.030001}      
  & This work  
  & BSE~\cite{Bhagwat_Maris:2008}    
  & Lattice QCD~\cite{Owen:2015}   
  & LFQM~\cite{Choi:1997, Choi:rho}
  & NJL model~\cite{Carrillo_NJL:2015, Ninomiya_NJL:2015}    
  \\[3pt]
  \hline \\[-6pt]
  %                                     %PDG           BLFQ       BSE            LATTICE         LFQM        NJL
  $\sqrt{\langle r^2_{\pi}\rangle}$     & 0.672(8)     & 0.44     & 0.66         & 0.591(43)     & 0.67      & 0.645 \\[4pt]
  $\sqrt{\langle r^2_{\rho}\rangle}$    & $-$          & 0.48     & 0.73         & 0.819(42)     & 0.52      & 0.82  \\[4pt]
  $\mu_\rho$                            & $-$          & 2.15     & 2.01         & 2.067(76)     & 1.92      & 2.48  \\[4pt]
  $\mathcal{Q}_\rho$                    & $-$          & -0.063   & -0.026       & -0.0452(61)   & -0.028    & -0.070 \\[5pt]
  \end{tabular}
\end{ruledtabular}
\label{tab:ff_obs}
\end{table*}

%%%%%%%%%%%%%%% SECTION 4 %%%%%%%%%%%%%%%%
\section{Parton Distribution Function and Parton Distribution Amplitude}
\subsection{Parton Distribution Function}
\begin{figure*}
  \centering
  \subfigure{\includegraphics[width=0.48\linewidth]{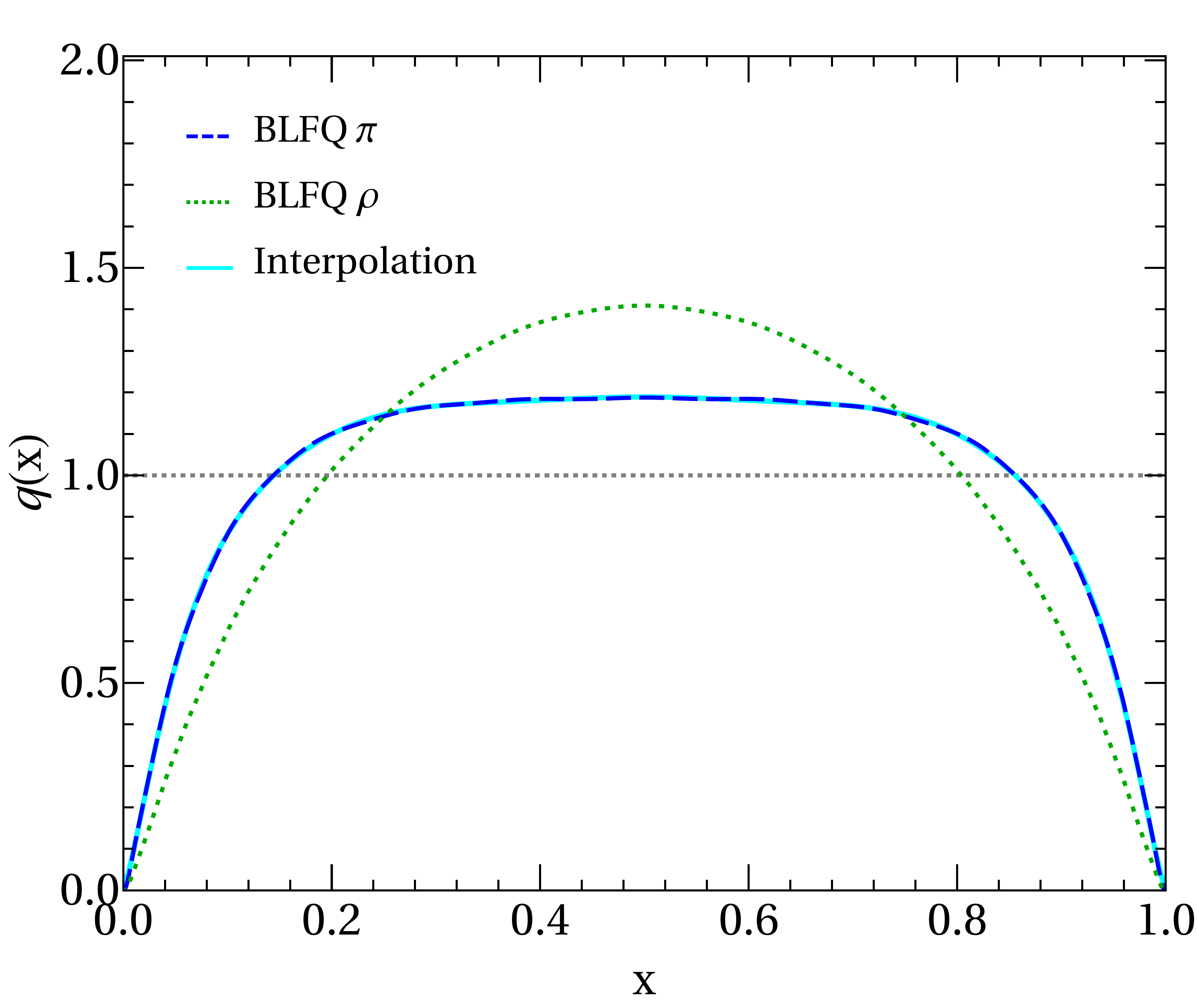}}\quad\quad
  \subfigure{\includegraphics[width=0.48\linewidth]{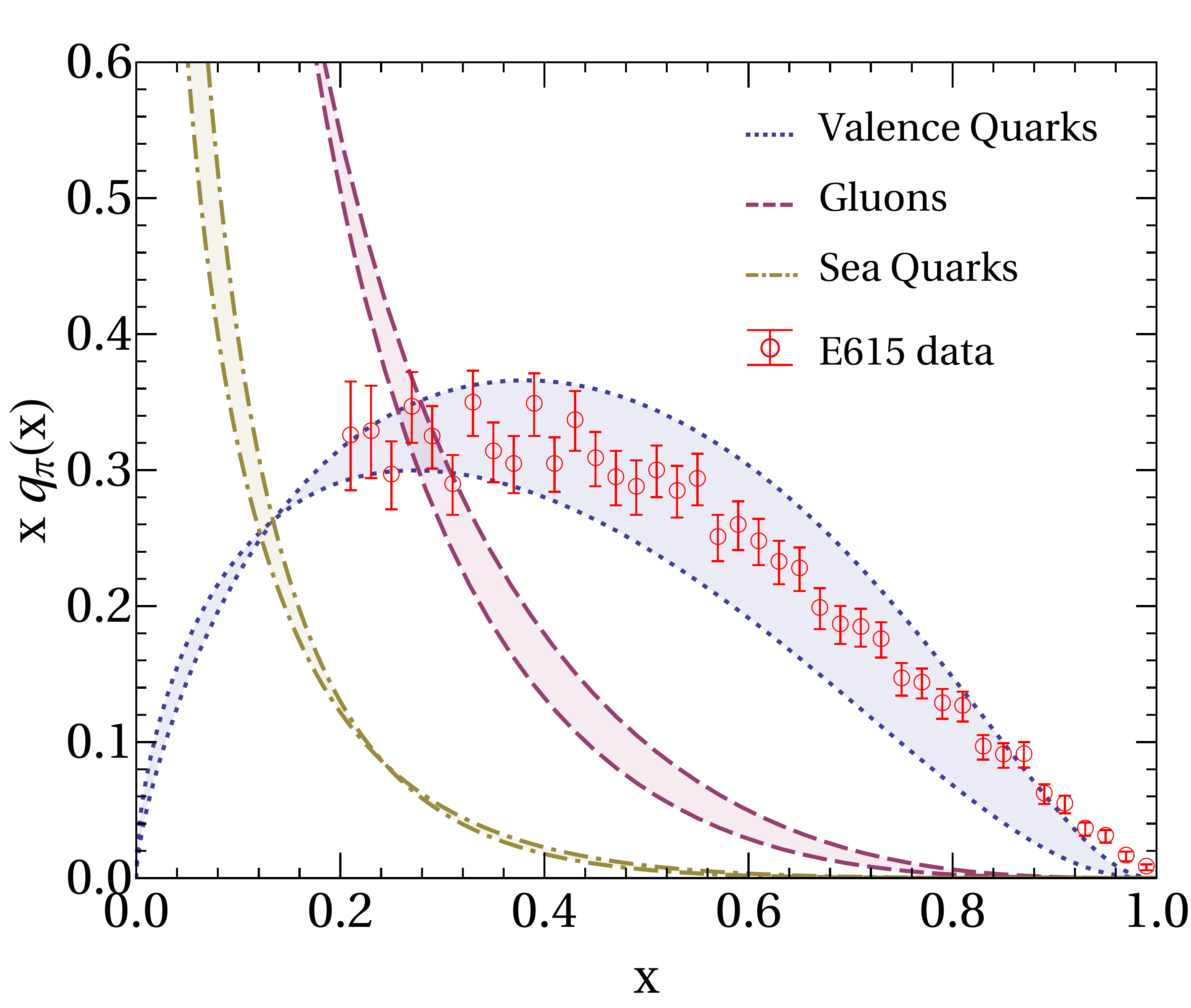}}
  \caption{\label{fig:pdf}BLFQ ($H_\text{eff, $\gamma_5$}$) calculations of PDFs for $\pi$ and $\rho$ at $N_\mathrm{max}=8$ and $L_\mathrm{max}=24$ (\textit{left panel}) according to Eq.~\eqref{eq:pdf} and the evolved pion PDF to the E615 experimental scale of 4.05 GeV using DGLAP evolution (\textit{right panel}). The polynomial interpolation of the pion PDF is used in the evolution with the HOPPET~\cite{Salam:2008qg}. Final valence quark results after the evolution are compared with the original analysis of the E615 experimental data~\cite{E615_Conway}, along with contributions from sea quarks and gluons.}
\end{figure*}

PDFs control the inclusive processes at large momentum transfer. The PDF is the probability density for finding a particle with certain longitudinal momentum fraction $x$ at a given factorization scale $\mu$. In the LFWF representation, it can be obtained by simply integrating out the transverse momentum of the squared wave function,
\begin{align}\label{eq:pdf}
q(x;\mu) = \frac{1}{4\pi x(1-x)}\sum_{s,\bar{s}} \int \frac{d^2\mathbf{k}_\perp}{(2\pi)^2}|\psi_{s,\bar{s}}(x,\mathbf{k}_\perp)|^2, 
\end{align}
where the scale indicated by $\mu$ is associated with the LFWF. Here the PDF and its first moment are normalized to 1, within the valence quark sectors as a consequence of the unit normalization of the valence LFWF given by Eq.~\eqref{eq:normalization}. In Fig. \ref{fig:pdf}, we present the valence PDFs for the pion and the rho meson at $N_\mathrm{max}=8$ and $L_\mathrm{max}=24$. In addition, we have also performed a polynomial interpolation of the pion PDF after factoring out the power-law end-point behavior.

The PDF for the $\pi$ obtained from the LFWF is expected to correspond to a low effective factorization scale compared to available experiments, as we only considered the valence quark sector and ignored contributions from sea quarks and gluons. We now evolve the PDFs from our model scale $\mu$ (a fit parameter described below) to the experimental scale via the well-known Dokshitzer-Gribov-Lipatov-Altarelli-Parisi (DGLAP) equations~\cite{Dokshitzer:1977sg, Gribov:1972ri, Altarelli:1977zs}, and compare the results at the experimental scale. Specifically, we evolve the parametric form of our initial pion PDF by using the next-to-next-to-leading order (NNLO) DGLAP with the Higher Order Perturbative Parton Evolution Toolkit (HOPPET)~\cite{Salam:2008qg}. The strong running coupling is specified by using the variable flavor-number scheme (VFNS) with the matching conditions at the heavy quark mass thresholds. It is initialized using the default value of $\alpha_{\mathrm{s}}(M_z^2)=0.1183$, and then carried out with the NNLO $\beta$-function. See Ref.~\cite{Salam:2008qg} for more details.
Since the initial scale of our model is estimated to be in the range between the IR regulator (0.21 GeV) and the UV regulator (1.83 GeV), we fit the evolved valence-quark PDF to the original Fermilab E615 data~\cite{E615_Conway} at its experimental scale of 4.05 GeV so as to determine our initial scale. In this way, the initial scale of our model is determined to be $\mu_0 =$ 0.56 GeV, which yields the results presented in Fig~\ref{fig:pdf}. We also include an error band showing the dependence of the pion PDF due to a 5\% assigned uncertainty in our initial energy scale. It is clear that the valence quark sectors are no longer the dominant contribution at this high scale (4.05 GeV); gluons and sea quarks become much more important especially in the small-$x$ region. At this scale, the valence quarks contribute around 43 \% of the hadron longitudinal momentum, whereas gluons and sea quarks contribute around 57 \% in total (43 \% gluon, 14 \% sea quarks).

\subsection{Parton Distribution Amplitude}
The PDAs control the exclusive processes at large momentum transfer~\cite{Lepage:1980}. In general, the PDAs are defined as the light-like separated gauge-invariant vacuum-to-meson matrix elements. Specifically in the LFWF formalism, PDAs can be expressed as~\cite{Li:2017mlw}
\begin{align}
\phi_{\mathrm{P,V}}(x; \mu) = \frac{2\sqrt{2N_c}}{f_{\mathrm{ P,V}}}\frac{1}{4\pi\sqrt{x(1-x)}} \int \frac{d^2 \mathbf{k}_\perp}{(2\pi)^2}\psi^{(m_j = 0)}_{\uparrow\downarrow \mp \downarrow\uparrow}(x, \mathbf{k}_\perp),\label{eq:pda}
\end{align}
where P and V stand for pseudoscalar and vector mesons, associated with the minus and plus signs respectively. Additionally, $f_{\mathrm{P,V}}$ are the decay constants defined in Eqs.~\eqref{eq:decay1} and \eqref{eq:decay2}, and $\mu$ is the factorization scale of the system, the dependence of which is given by the Efremov-Radyushkin-Brodsky-Lepage (ERBL) evolution~\cite{Lepage:1980, Efremov:1980, Efremov:1980_v2}. With these definitions, the PDAs at any scale are normalized to 1.

\begin{figure*}
  \centering
  \subfigure{\includegraphics[width=0.48\linewidth]{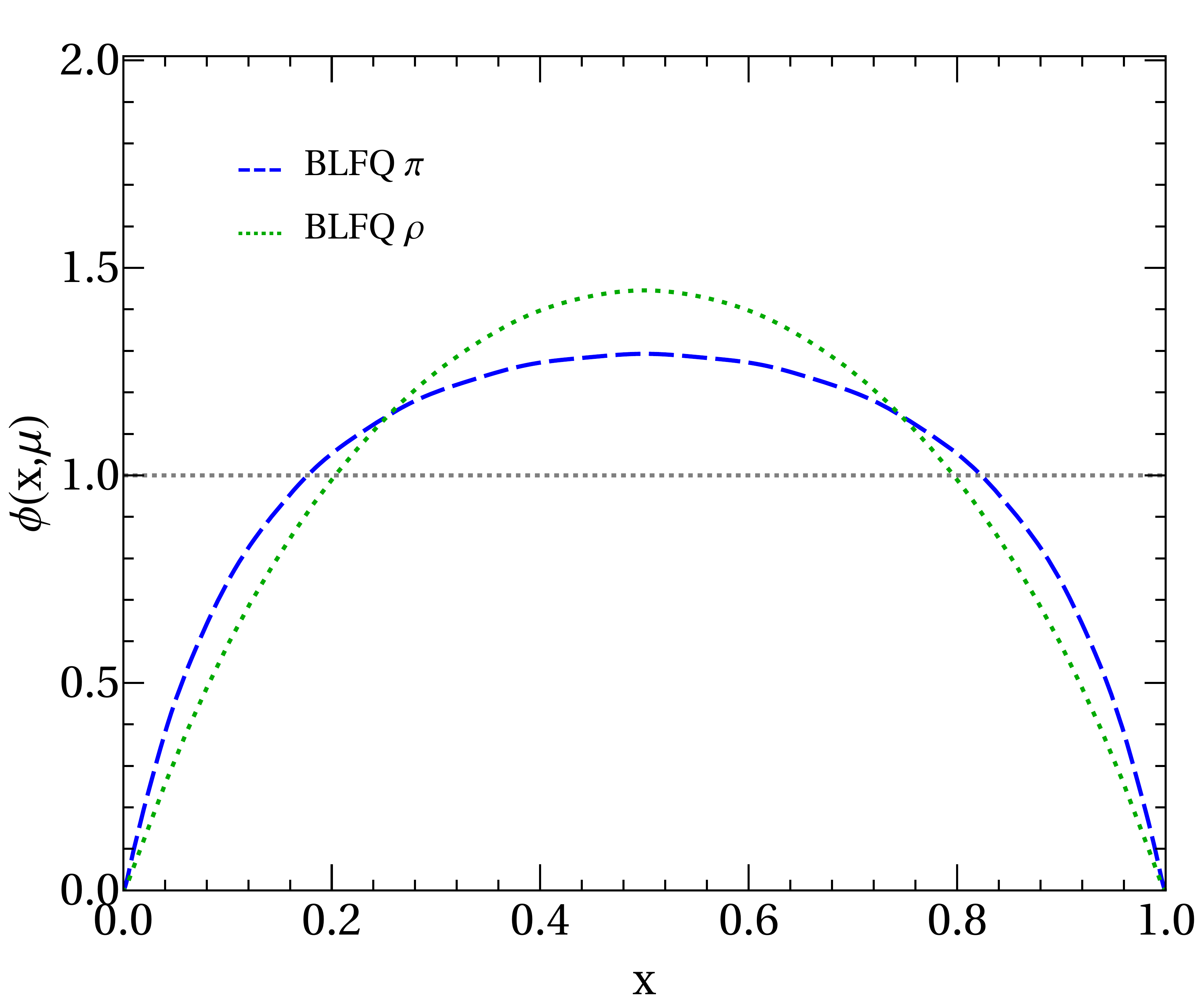}}\quad\quad
  \subfigure{\includegraphics[width=0.48\linewidth]{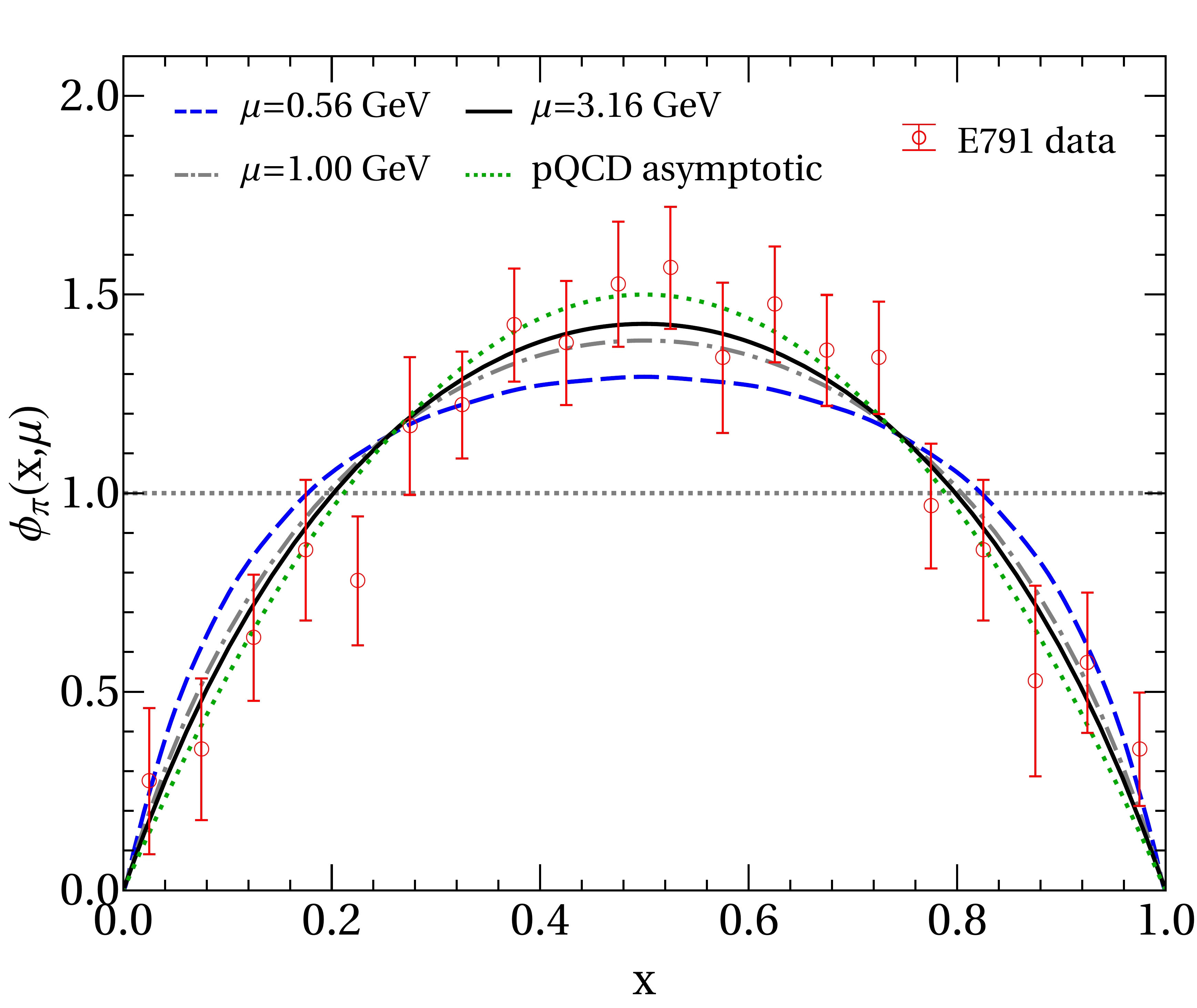}}
  \caption{\label{fig:pda}BLFQ ($H_\text{eff, $\gamma_5$}$) calculations of PDAs for $\pi$ and $\rho$ at $N_\mathrm{max}=8$ and $L_\mathrm{max}=24$ in the initial scale (\textit{left panel}) according to Eq.~\eqref{eq:pda} and the ERBL evolution of the pion PDA to the E791 experimental scale at 3.16 GeV~\cite{E791} (\textit{right panel}) using Eq.~\eqref{eq:erbl}. The initial PDA of the $\pi$ and that of the $\rho$ are shown on the left as the dashed line and the dotted line, respectively. On the right, the evolutions of the pion PDA at various energy scales $\mu$ = 0.56 GeV (dashed line), 1.00 GeV (dot-dashed line), 3.16 GeV (solid line) are presented. The asymptotic PDA is included as the dotted line for comparison.}
\end{figure*}

By assuming the PDAs having the same initial scale as that of the pion PDF, we present the PDAs for the pion and the rho meson at 0.56 GeV in Fig~\ref{fig:pda}. One should notice that the PDA for the $\pi$ is much broader than that of the $\rho$ at this low energy scale, which suggests that the $\pi$ is a more relativistic system than the $\rho$. In order to compare the pion PDA with the available experimental data from Fermilab E791~\cite{E791}, we performed the ERBL evolution~\cite{Ruiz} of this initial pion PDA.

In the Gegenbauer basis, the evolved PDA of the pion at any scale $\mu$ is given by
\begin{align}
  \phi_{\pi}(x, \mu) = 6 x (1-x) \sum_{n=0}^{\infty} C_n^{3/2} (2x-1) \ a_n(\mu, \mu_0),\label{eq:erbl}
\end{align}
where 
\begin{align}
  a_n(\mu, \mu_0) &= \frac{2}{3}\frac{2n+3}{(n+1)(n+2)}\bigg (\frac{\alpha_{\mathrm{s}}(\mu)}{\alpha_{\mathrm{s}}(\mu_0)}\bigg)^{4\pi\gamma^0_n/2\beta_0} 
  \nonumber\\
  &\quad\times\int_0^1 dx \ C_n^{3/2}     \ (2x-1) \ \phi_{\pi}(x, \mu_0).
\end{align}
Here $C_n^{3/2} (2x-1)$ is the Gegenbauer polynomial, $\mu_0$ is the initial scale, and $\gamma^0_n = -2 \ C_{\mathrm{F}} \ \big ( 3 + \frac{2}{(n+1)(n+2)} - 4 \sum_{k=1}^{n+1} \frac{1}{k} \big)$. In Fig.~\ref{fig:pda}, we present the results of the ERBL evolution of the pion PDA at various energy scales using the same running coupling as in the QCD evolution of the pion PDF. As the energy increases from 0.54 GeV to 3.16 GeV (E791 experimental energy), the pion PDA gradually trends towards the perturbative QCD asymptotic PDA, $\phi(x, \infty)=6x(1-x)$. It is interesting to point out that our BLFQ model of the pion begins its evolution below the pQCD asymptotic at $x=0.5$, which is opposite to some of the available calculations, including AdS/QCD~\cite{Ahmady:dynamical_spin}.

We have also calculated $n$-th moment of the PDA, $\langle \xi^n \rangle = \int_0^1 dx(2x-1)^n \phi(x)$, in order to compare with other approaches quantitatively. In Table~\ref{tab:moments}, we compare our moment calculations for the $\pi$ and $\rho$ with other available models. Notice that only the $n$-th moments with even $n$ are shown here as any odd power vanishes due to isospin symmetry. The second moment can also be used to estimate the relative velocity of the parton via $\langle v^2 \rangle \approx 3 \langle \xi^2 \rangle$. Here, $\langle v^2_{\pi} \rangle \approx 0.66 $ and $\langle v^2_{\rho} \rangle \approx 0.61 $ at 1 GeV, much higher than those of heavy quarkonia~\cite{Li:2017mlw}, where $\langle v^2_{\eta_c} \rangle \approx 0.36 $ and $\langle v^2_{\eta_b} \rangle \approx 0.21 $.

\begin{table*}[!htbp]
\centering
\caption{\label{tab:moments}Leading moments of $\pi$ meson and $\rho$ meson calculated using their PDAs in BLFQ ($H_\text{eff, $\gamma_5$}$) at $N_\mathrm{max}=8$ and $L_\mathrm{max}=24$, compared with selected results obtained from AdS/QCD, BSE, pQCD, QCD sum rules (SR), lattice QCD, and light-front quark model (LFQM). Note that results from various models are obtained at similiar but slightly different energy scales $\mu$ estimated by the corresponding references. We show results of both rainbow-ladder truncation (left) and improved kernels (right) for the BSE approach, and results of both harmonic oscillator (left) and linear (right) confining potentials for the LFQM approach. The numerical uncertainty of our result is under control by using Gauss-Legendre quadrature rule with a sufficient number of Gaussian weights and precisions. Uncertainties from other approaches, where available, are quoted in parenthesis when they are symmetrical or explicitly when they are asymmetrical.} 
\begin{ruledtabular}
  \begin{tabular}{ c  c@{\hskip 0.3in} c@{\hskip 0.3in}  c  c  c  c  c  c@{\hskip 0.13in}  c}
  \\[-9pt]
  & $\pi$
  & This work 
  & pQCD 
  & AdS/QCD~\cite{holography_pion}  
  & BSE~\cite{DSE_Chang}   
  & SR~\cite{Stefanis:2015qha}
  & SR~\cite{Ball_sr}
  & Lattice QCD~\cite{Arthur:light_meson} 
  & LFQM~\cite{Choi:pion_rho} 
  \\[7pt] \hline 
  \\[-6pt]
  \multirow{3}{*}{} 
                           % BLFQ         % pQCD    % AdS     % BSE              % SR                           % SR         % Lattice     % LFQM
  & $\mu \text{ (GeV)}$    & 1, 3.16       &$\infty$ & 1       & 2               & 2                            & 1          & 2           & 1           \\[4pt] 
  & $\langle \xi^2\rangle$ & 0.220, 0.213  & 0.200   & 0.237   & 0.280, 0.251    & $0.220_{-0.006}^{+0.009}$    & 0.24       & 0.28(1)(2)  & 0.24, 0.22  \\[4pt]
  & $\langle \xi^4\rangle$ & 0.099, 0.095  & 0.086   & 0.114   & 0.151, 0.128    & $0.098_{-0.005}^{+0.008}$    & 0.11       &             & 0.11, 0.09  \\[4pt]
  & $\langle \xi^6\rangle$ & 0.057, 0.054  & 0.048   & 0.078   & 0.099, 0.081    &                              &            &             & 0.07, 0.05  \\[9pt] 
  \hline \hline
  \\[-6pt]
  & $\rho$
  & This work 
  & pQCD 
  & AdS/QCD~\cite{Forshaw_rho_meson}  
  & BSE~\cite{Gao_bse}   
  & SR~\cite{Stefanis:2015qha} 
  & SR~\cite{Bakulev_rho} 
  & Lattice QCD~\cite{Arthur:light_meson} 
  & LFQM~\cite{Choi:pion_rho} 
  \\[7pt] \hline 
  \\[-6pt] 
  \multirow{3}{*}{} 
                           % BLFQ          % pQCD    % AdS     % BSE        % SR        % SR         % Lattice     % LFQM
  & $\mu \text{ (GeV)}$    & 1, 2          &$\infty$ & 1       & 2          & 2         & 1          & 2           & 1               \\[4pt] 
  & $\langle \xi^2\rangle$ & 0.204, 0.203  & 0.200   & 0.227   & 0.23       & 0.206(8)  & 0.227(7)   & 0.27(1)(2)  & 0.21, 0.19      \\[4pt]
  & $\langle \xi^4\rangle$ & 0.089, 0.088  & 0.086   & 0.105   & 0.11       & 0.087(6)  & 0.095(5)   &             & 0.09, 0.08      \\[4pt]
  & $\langle \xi^6\rangle$ & 0.050, 0.049  & 0.048   & 0.062   & 0.066      &           & 0.051(4)   &             & 0.05, 0.04      \\[6pt] 
  \end{tabular}
\end{ruledtabular}
\end{table*}

\section{Summary and Discussions}\label{sec:summary}
In this work, we investigated the light unflavored mesons within the BLFQ approach. Our model Hamiltonian was extended from the effective Hamiltonian used for heavy quarkonia by adding the pseudoscalar contact interaction inspired by the NJL model. Within our two-body basis representation, we solved for the mass spectrum and the LFWFs. We fitted three model parameters for the light-unflavored mesons, the quark mass, the confining strength, and the contact interaction coupling, using the meson masses given by the PDG. The resulting spectroscopy agreed with 11 states in the PDG to within a r.m.s. mass deviation of 111 MeV. We showed that the inclusion of pseudoscalar contact interaction was valuable to account for the $\pi$-$\rho$ mass splitting.

In addition, we studied the internal structures of the pion and the rho meson by analyzing their LFWFs and subsequent observables. With these LFWFs, we calculated decay constants for the $\pi$ and the $\rho$ mesons. Furthermore, we computed their electromagnetic form factors and subsequently obtained physical observables including the charge radii, the magnetic moment, and the quadrupole moment. We attributed the small charge radius of the pion to the use of point-like valence quarks in our models. We also calculated the PDFs and PDAs for the $\pi$ and the $\rho$ at the model scale, and used DGLAP and ERBL evolutions respectively to evaluate the PDF and PDA at experimental scales for the $\pi$. Taking the initial scale as a fit parameter, the results for the $\pi$ were in good agreement with available data and other models. PDFs after the DGLAP evolution also illustrated the significant roles taken by the gluon and the sea quarks in light mesons at the experimental scale.

This work is our first step to understand light mesons within the light-front Hamiltonian formalism. We expect to include self-energy corrections and higher Fock sectors for a more comprehensive description of the light meson systems. We anticipate that inclusion of these additional degrees of freedom and interactions in the Hamiltonian will improve the current results towards a better agreement with experiment and more fundamental treatment of dynamical chiral symmetry breaking.

\section*{Acknowledgements}
We wish to thank M. Li, C. Mondal, X. Zhao, P. Maris,  J. Lan, S. Tang, and A. Trawi\'{n}ski for valuable discussions. We would also like to thank W. Broniowski for kindly providing us the experimental data for the pion PDA from FNAL-E-0791. This work was supported in part by the US Department of Energy (DOE) under Grant Nos. DE-FG02-87ER40371, DESC0018223 (SciDAC-4/NUCLEI), DE-SC0015376 (DOE Topical Collaboration in Nuclear Theory for Double-Beta Decay and Fundamental Symmetries). This research used resources of the National Energy Research Scientific Computing Center (NERSC), a U.S. Department of Energy Office of Science User Facility operated under Contract No. DE-AC02-05CH11231.

\bibliography{bibliography}

\begin{thebibliography}{65}%
\makeatletter
\providecommand \@ifxundefined [1]{%
 \@ifx{#1\undefined}
}%
\providecommand \@ifnum [1]{%
 \ifnum #1\expandafter \@firstoftwo
 \else \expandafter \@secondoftwo
 \fi
}%
\providecommand \@ifx [1]{%
 \ifx #1\expandafter \@firstoftwo
 \else \expandafter \@secondoftwo
 \fi
}%
\providecommand \natexlab [1]{#1}%
\providecommand \enquote  [1]{``#1''}%
\providecommand \bibnamefont  [1]{#1}%
\providecommand \bibfnamefont [1]{#1}%
\providecommand \citenamefont [1]{#1}%
\providecommand \href@noop [0]{\@secondoftwo}%
\providecommand \href [0]{\begingroup \@sanitize@url \@href}%
\providecommand \@href[1]{\@@startlink{#1}\@@href}%
\providecommand \@@href[1]{\endgroup#1\@@endlink}%
\providecommand \@sanitize@url [0]{\catcode `\\12\catcode `\$12\catcode
  `\&12\catcode `\#12\catcode `\^12\catcode `\_12\catcode `\%12\relax}%
\providecommand \@@startlink[1]{}%
\providecommand \@@endlink[0]{}%
\providecommand \url  [0]{\begingroup\@sanitize@url \@url }%
\providecommand \@url [1]{\endgroup\@href {#1}{\urlprefix }}%
\providecommand \urlprefix  [0]{URL }%
\providecommand \Eprint [0]{\href }%
\providecommand \doibase [0]{http://dx.doi.org/}%
\providecommand \selectlanguage [0]{\@gobble}%
\providecommand \bibinfo  [0]{\@secondoftwo}%
\providecommand \bibfield  [0]{\@secondoftwo}%
\providecommand \translation [1]{[#1]}%
\providecommand \BibitemOpen [0]{}%
\providecommand \bibitemStop [0]{}%
\providecommand \bibitemNoStop [0]{.\EOS\space}%
\providecommand \EOS [0]{\spacefactor3000\relax}%
\providecommand \BibitemShut  [1]{\csname bibitem#1\endcsname}%
\let\auto@bib@innerbib\@empty
%</preamble>
\bibitem [{\citenamefont {Maris}\ and\ \citenamefont {Tandy}(1999)}]{Maris}%
  \BibitemOpen
  \bibfield  {author} {\bibinfo {author} {\bibfnamefont {P.}~\bibnamefont
  {Maris}}\ and\ \bibinfo {author} {\bibfnamefont {P.~C.}\ \bibnamefont
  {Tandy}},\ }\href {\doibase 10.1103/PhysRevC.60.055214} {\bibfield  {journal}
  {\bibinfo  {journal} {Phys. Rev.}\ }\textbf {\bibinfo {volume} {C60}},\
  \bibinfo {pages} {055214} (\bibinfo {year} {1999})},\ \Eprint
  {http://arxiv.org/abs/nucl-th/9905056} {arXiv:nucl-th/9905056 [nucl-th]}
  \BibitemShut {NoStop}%
%%CITATION = NUCL-TH/9905056;%%
\bibitem [{\citenamefont {Chang}\ \emph {et~al.}(2013)\citenamefont {Chang},
  \citenamefont {Cloet}, \citenamefont {Cobos-Martinez}, \citenamefont
  {Roberts}, \citenamefont {Schmidt},\ and\ \citenamefont {Tandy}}]{DSE_Chang}%
  \BibitemOpen
  \bibfield  {author} {\bibinfo {author} {\bibfnamefont {L.}~\bibnamefont
  {Chang}}, \bibinfo {author} {\bibfnamefont {I.~C.}\ \bibnamefont {Cloet}},
  \bibinfo {author} {\bibfnamefont {J.~J.}\ \bibnamefont {Cobos-Martinez}},
  \bibinfo {author} {\bibfnamefont {C.~D.}\ \bibnamefont {Roberts}}, \bibinfo
  {author} {\bibfnamefont {S.~M.}\ \bibnamefont {Schmidt}}, \ and\ \bibinfo
  {author} {\bibfnamefont {P.~C.}\ \bibnamefont {Tandy}},\ }\href {\doibase
  10.1103/PhysRevLett.110.132001} {\bibfield  {journal} {\bibinfo  {journal}
  {Phys. Rev. Lett.}\ }\textbf {\bibinfo {volume} {110}},\ \bibinfo {pages}
  {132001} (\bibinfo {year} {2013})},\ \Eprint {http://arxiv.org/abs/1301.0324}
  {arXiv:1301.0324 [nucl-th]} \BibitemShut {NoStop}%
%%CITATION = ARXIV:1301.0324;%%
\bibitem [{\citenamefont {Fischer}\ \emph {et~al.}(2014)\citenamefont
  {Fischer}, \citenamefont {Kubrak},\ and\ \citenamefont
  {Williams}}]{Fischer:2014xha}%
  \BibitemOpen
  \bibfield  {author} {\bibinfo {author} {\bibfnamefont {C.~S.}\ \bibnamefont
  {Fischer}}, \bibinfo {author} {\bibfnamefont {S.}~\bibnamefont {Kubrak}}, \
  and\ \bibinfo {author} {\bibfnamefont {R.}~\bibnamefont {Williams}},\ }\href
  {\doibase 10.1140/epja/i2014-14126-6} {\bibfield  {journal} {\bibinfo
  {journal} {Eur. Phys. J.}\ }\textbf {\bibinfo {volume} {A50}},\ \bibinfo
  {pages} {126} (\bibinfo {year} {2014})},\ \Eprint
  {http://arxiv.org/abs/1406.4370} {arXiv:1406.4370 [hep-ph]} \BibitemShut
  {NoStop}%
%%CITATION = ARXIV:1406.4370;%%
\bibitem [{\citenamefont {Williams}\ \emph {et~al.}(2016)\citenamefont
  {Williams}, \citenamefont {Fischer},\ and\ \citenamefont
  {Heupel}}]{Williams:2015cvx}%
  \BibitemOpen
  \bibfield  {author} {\bibinfo {author} {\bibfnamefont {R.}~\bibnamefont
  {Williams}}, \bibinfo {author} {\bibfnamefont {C.~S.}\ \bibnamefont
  {Fischer}}, \ and\ \bibinfo {author} {\bibfnamefont {W.}~\bibnamefont
  {Heupel}},\ }\href {\doibase 10.1103/PhysRevD.93.034026} {\bibfield
  {journal} {\bibinfo  {journal} {Phys. Rev.}\ }\textbf {\bibinfo {volume}
  {D93}},\ \bibinfo {pages} {034026} (\bibinfo {year} {2016})},\ \Eprint
  {http://arxiv.org/abs/1512.00455} {arXiv:1512.00455 [hep-ph]} \BibitemShut
  {NoStop}%
%%CITATION = ARXIV:1512.00455;%%
\bibitem [{\citenamefont {Braun}\ \emph {et~al.}(2006)\citenamefont {Braun},
  \citenamefont {G\"ockeler}, \citenamefont {Horsley}, \citenamefont {Perlt},
  \citenamefont {Pleiter}, \citenamefont {Rakow}, \citenamefont {Schierholz},
  \citenamefont {Schiller}, \citenamefont {Schroers}, \citenamefont
  {St\"uben},\ and\ \citenamefont {Zanotti}}]{Braun:pion_moments}%
  \BibitemOpen
  \bibfield  {author} {\bibinfo {author} {\bibfnamefont {V.~M.}\ \bibnamefont
  {Braun}}, \bibinfo {author} {\bibfnamefont {M.}~\bibnamefont {G\"ockeler}},
  \bibinfo {author} {\bibfnamefont {R.}~\bibnamefont {Horsley}}, \bibinfo
  {author} {\bibfnamefont {H.}~\bibnamefont {Perlt}}, \bibinfo {author}
  {\bibfnamefont {D.}~\bibnamefont {Pleiter}}, \bibinfo {author} {\bibfnamefont
  {P.~E.~L.}\ \bibnamefont {Rakow}}, \bibinfo {author} {\bibfnamefont
  {G.}~\bibnamefont {Schierholz}}, \bibinfo {author} {\bibfnamefont
  {A.}~\bibnamefont {Schiller}}, \bibinfo {author} {\bibfnamefont
  {W.}~\bibnamefont {Schroers}}, \bibinfo {author} {\bibfnamefont
  {H.}~\bibnamefont {St\"uben}}, \ and\ \bibinfo {author} {\bibfnamefont
  {J.~M.}\ \bibnamefont {Zanotti}},\ }\href {\doibase
  10.1103/PhysRevD.74.074501} {\bibfield  {journal} {\bibinfo  {journal} {Phys.
  Rev.}\ }\textbf {\bibinfo {volume} {D74}},\ \bibinfo {pages} {074501}
  (\bibinfo {year} {2006})},\ \Eprint {http://arxiv.org/abs/hep-lat/0606012}
  {arXiv:hep-lat/0606012 [hep-lat]} \BibitemShut {NoStop}%
%%CITATION = HEP-LAT/0606012;%%
\bibitem [{\citenamefont {Braun}\ \emph {et~al.}(2015)\citenamefont {Braun},
  \citenamefont {Collins}, \citenamefont {Göckeler}, \citenamefont
  {Pérez-Rubio}, \citenamefont {Schäfer}, \citenamefont {Schiel},\ and\
  \citenamefont {Sternbeck}}]{Braun_v2}%
  \BibitemOpen
  \bibfield  {author} {\bibinfo {author} {\bibfnamefont {V.~M.}\ \bibnamefont
  {Braun}}, \bibinfo {author} {\bibfnamefont {S.}~\bibnamefont {Collins}},
  \bibinfo {author} {\bibfnamefont {M.}~\bibnamefont {Göckeler}}, \bibinfo
  {author} {\bibfnamefont {P.}~\bibnamefont {Pérez-Rubio}}, \bibinfo {author}
  {\bibfnamefont {A.}~\bibnamefont {Schäfer}}, \bibinfo {author}
  {\bibfnamefont {R.~W.}\ \bibnamefont {Schiel}}, \ and\ \bibinfo {author}
  {\bibfnamefont {A.}~\bibnamefont {Sternbeck}},\ }\href {\doibase
  10.1103/PhysRevD.92.014504} {\bibfield  {journal} {\bibinfo  {journal} {Phys.
  Rev.}\ }\textbf {\bibinfo {volume} {D92}},\ \bibinfo {pages} {014504}
  (\bibinfo {year} {2015})},\ \Eprint {http://arxiv.org/abs/1503.03656}
  {arXiv:1503.03656 [hep-lat]} \BibitemShut {NoStop}%
%%CITATION = ARXIV:1503.03656;%%
\bibitem [{\citenamefont {Arthur}\ \emph {et~al.}(2011)\citenamefont {Arthur},
  \citenamefont {Boyle}, \citenamefont {Brommel}, \citenamefont {Donnellan},
  \citenamefont {Flynn}, \citenamefont {Juttner}, \citenamefont {Rae},\ and\
  \citenamefont {Sachrajda}}]{Arthur:light_meson}%
  \BibitemOpen
  \bibfield  {author} {\bibinfo {author} {\bibfnamefont {R.}~\bibnamefont
  {Arthur}}, \bibinfo {author} {\bibfnamefont {P.~A.}\ \bibnamefont {Boyle}},
  \bibinfo {author} {\bibfnamefont {D.}~\bibnamefont {Brommel}}, \bibinfo
  {author} {\bibfnamefont {M.~A.}\ \bibnamefont {Donnellan}}, \bibinfo {author}
  {\bibfnamefont {J.~M.}\ \bibnamefont {Flynn}}, \bibinfo {author}
  {\bibfnamefont {A.}~\bibnamefont {Juttner}}, \bibinfo {author} {\bibfnamefont
  {T.~D.}\ \bibnamefont {Rae}}, \ and\ \bibinfo {author} {\bibfnamefont
  {C.~T.~C.}\ \bibnamefont {Sachrajda}},\ }\href {\doibase
  10.1103/PhysRevD.83.074505} {\bibfield  {journal} {\bibinfo  {journal} {Phys.
  Rev.}\ }\textbf {\bibinfo {volume} {D83}},\ \bibinfo {pages} {074505}
  (\bibinfo {year} {2011})},\ \Eprint {http://arxiv.org/abs/1011.5906}
  {arXiv:1011.5906 [hep-lat]} \BibitemShut {NoStop}%
%%CITATION = ARXIV:1011.5906;%%
\bibitem [{\citenamefont {Sufian}\ \emph {et~al.}(2019)\citenamefont {Sufian},
  \citenamefont {Karpie}, \citenamefont {Egerer}, \citenamefont {Orginos},
  \citenamefont {Qiu},\ and\ \citenamefont {Richards}}]{Sufian}%
  \BibitemOpen
  \bibfield  {author} {\bibinfo {author} {\bibfnamefont {R.~S.}\ \bibnamefont
  {Sufian}}, \bibinfo {author} {\bibfnamefont {J.}~\bibnamefont {Karpie}},
  \bibinfo {author} {\bibfnamefont {C.}~\bibnamefont {Egerer}}, \bibinfo
  {author} {\bibfnamefont {K.}~\bibnamefont {Orginos}}, \bibinfo {author}
  {\bibfnamefont {J.-W.}\ \bibnamefont {Qiu}}, \ and\ \bibinfo {author}
  {\bibfnamefont {D.~G.}\ \bibnamefont {Richards}},\ }\href {\doibase
  10.1103/PhysRevD.99.074507} {\bibfield  {journal} {\bibinfo  {journal} {Phys.
  Rev.}\ }\textbf {\bibinfo {volume} {D99}},\ \bibinfo {pages} {074507}
  (\bibinfo {year} {2019})},\ \Eprint {http://arxiv.org/abs/1901.03921}
  {arXiv:1901.03921 [hep-lat]} \BibitemShut {NoStop}%
%%CITATION = ARXIV:1901.03921;%%
\bibitem [{\citenamefont {Cloët}\ \emph {et~al.}(2013)\citenamefont {Cloët},
  \citenamefont {Chang}, \citenamefont {Roberts}, \citenamefont {Schmidt},\
  and\ \citenamefont {Tandy}}]{Cloet_lattice}%
  \BibitemOpen
  \bibfield  {author} {\bibinfo {author} {\bibfnamefont {I.~C.}\ \bibnamefont
  {Cloët}}, \bibinfo {author} {\bibfnamefont {L.}~\bibnamefont {Chang}},
  \bibinfo {author} {\bibfnamefont {C.~D.}\ \bibnamefont {Roberts}}, \bibinfo
  {author} {\bibfnamefont {S.~M.}\ \bibnamefont {Schmidt}}, \ and\ \bibinfo
  {author} {\bibfnamefont {P.~C.}\ \bibnamefont {Tandy}},\ }\href {\doibase
  10.1103/PhysRevLett.111.092001} {\bibfield  {journal} {\bibinfo  {journal}
  {Phys. Rev. Lett.}\ }\textbf {\bibinfo {volume} {111}},\ \bibinfo {pages}
  {092001} (\bibinfo {year} {2013})},\ \Eprint {http://arxiv.org/abs/1306.2645}
  {arXiv:1306.2645 [nucl-th]} \BibitemShut {NoStop}%
%%CITATION = ARXIV:1306.2645;%%
\bibitem [{\citenamefont {Segovia}\ \emph {et~al.}(2014)\citenamefont
  {Segovia}, \citenamefont {Chang}, \citenamefont {Cloët}, \citenamefont
  {Roberts}, \citenamefont {Schmidt},\ and\ \citenamefont
  {Zong}}]{Segovia:2013}%
  \BibitemOpen
  \bibfield  {author} {\bibinfo {author} {\bibfnamefont {J.}~\bibnamefont
  {Segovia}}, \bibinfo {author} {\bibfnamefont {L.}~\bibnamefont {Chang}},
  \bibinfo {author} {\bibfnamefont {I.~C.}\ \bibnamefont {Cloët}}, \bibinfo
  {author} {\bibfnamefont {C.~D.}\ \bibnamefont {Roberts}}, \bibinfo {author}
  {\bibfnamefont {S.~M.}\ \bibnamefont {Schmidt}}, \ and\ \bibinfo {author}
  {\bibfnamefont {H.-s.}\ \bibnamefont {Zong}},\ }\href {\doibase
  10.1016/j.physletb.2014.02.006} {\bibfield  {journal} {\bibinfo  {journal}
  {Phys. Lett.}\ }\textbf {\bibinfo {volume} {B731}},\ \bibinfo {pages} {13}
  (\bibinfo {year} {2014})},\ \Eprint {http://arxiv.org/abs/1311.1390}
  {arXiv:1311.1390 [nucl-th]} \BibitemShut {NoStop}%
%%CITATION = ARXIV:1311.1390;%%
\bibitem [{\citenamefont {Brodsky}\ \emph {et~al.}(2015)\citenamefont
  {Brodsky}, \citenamefont {de~Teramond}, \citenamefont {Dosch},\ and\
  \citenamefont {Erlich}}]{holography}%
  \BibitemOpen
  \bibfield  {author} {\bibinfo {author} {\bibfnamefont {S.~J.}\ \bibnamefont
  {Brodsky}}, \bibinfo {author} {\bibfnamefont {G.~F.}\ \bibnamefont
  {de~Teramond}}, \bibinfo {author} {\bibfnamefont {H.~G.}\ \bibnamefont
  {Dosch}}, \ and\ \bibinfo {author} {\bibfnamefont {J.}~\bibnamefont
  {Erlich}},\ }\href {\doibase 10.1016/j.physrep.2015.05.001} {\bibfield
  {journal} {\bibinfo  {journal} {Phys. Rept.}\ }\textbf {\bibinfo {volume}
  {584}},\ \bibinfo {pages} {1} (\bibinfo {year} {2015})},\ \Eprint
  {http://arxiv.org/abs/1407.8131} {arXiv:1407.8131 [hep-ph]} \BibitemShut
  {NoStop}%
%%CITATION = ARXIV:1407.8131;%%
\bibitem [{\citenamefont {Ahmady}\ \emph {et~al.}(2018)\citenamefont {Ahmady},
  \citenamefont {Mondal},\ and\ \citenamefont
  {Sandapen}}]{Ahmady:dynamical_spin}%
  \BibitemOpen
  \bibfield  {author} {\bibinfo {author} {\bibfnamefont {M.}~\bibnamefont
  {Ahmady}}, \bibinfo {author} {\bibfnamefont {C.}~\bibnamefont {Mondal}}, \
  and\ \bibinfo {author} {\bibfnamefont {R.}~\bibnamefont {Sandapen}},\ }\href
  {\doibase 10.1103/PhysRevD.98.034010} {\bibfield  {journal} {\bibinfo
  {journal} {Phys. Rev.}\ }\textbf {\bibinfo {volume} {D98}},\ \bibinfo {pages}
  {034010} (\bibinfo {year} {2018})},\ \Eprint
  {http://arxiv.org/abs/1805.08911} {arXiv:1805.08911 [hep-ph]} \BibitemShut
  {NoStop}%
%%CITATION = ARXIV:1805.08911;%%
\bibitem [{\citenamefont {Choi}\ and\ \citenamefont
  {Ji}(2007)}]{Choi:pion_rho}%
  \BibitemOpen
  \bibfield  {author} {\bibinfo {author} {\bibfnamefont {H.-M.}\ \bibnamefont
  {Choi}}\ and\ \bibinfo {author} {\bibfnamefont {C.-R.}\ \bibnamefont {Ji}},\
  }\href {\doibase 10.1103/PhysRevD.75.034019} {\bibfield  {journal} {\bibinfo
  {journal} {Phys. Rev.}\ }\textbf {\bibinfo {volume} {D75}},\ \bibinfo {pages}
  {034019} (\bibinfo {year} {2007})},\ \Eprint
  {http://arxiv.org/abs/hep-ph/0701177} {arXiv:hep-ph/0701177 [hep-ph]}
  \BibitemShut {NoStop}%
%%CITATION = HEP-PH/0701177;%%
\bibitem [{\citenamefont {Choi}\ and\ \citenamefont {Ji}(2015)}]{Choi:2015}%
  \BibitemOpen
  \bibfield  {author} {\bibinfo {author} {\bibfnamefont {H.-M.}\ \bibnamefont
  {Choi}}\ and\ \bibinfo {author} {\bibfnamefont {C.-R.}\ \bibnamefont {Ji}},\
  }\href {\doibase 10.1103/PhysRevD.91.014018} {\bibfield  {journal} {\bibinfo
  {journal} {Phys. Rev.}\ }\textbf {\bibinfo {volume} {D91}},\ \bibinfo {pages}
  {014018} (\bibinfo {year} {2015})},\ \Eprint {http://arxiv.org/abs/1412.2507}
  {arXiv:1412.2507 [hep-ph]} \BibitemShut {NoStop}%
%%CITATION = ARXIV:1412.2507;%%
\bibitem [{\citenamefont {Choi}\ and\ \citenamefont {Ji}(1999)}]{Choi:1997}%
  \BibitemOpen
  \bibfield  {author} {\bibinfo {author} {\bibfnamefont {H.-M.}\ \bibnamefont
  {Choi}}\ and\ \bibinfo {author} {\bibfnamefont {C.-R.}\ \bibnamefont {Ji}},\
  }\href {\doibase 10.1103/PhysRevD.59.074015} {\bibfield  {journal} {\bibinfo
  {journal} {Phys. Rev.}\ }\textbf {\bibinfo {volume} {D59}},\ \bibinfo {pages}
  {074015} (\bibinfo {year} {1999})},\ \Eprint
  {http://arxiv.org/abs/hep-ph/9711450} {arXiv:hep-ph/9711450 [hep-ph]}
  \BibitemShut {NoStop}%
%%CITATION = HEP-PH/9711450;%%
\bibitem [{\citenamefont {Spence}\ and\ \citenamefont
  {Vary}(1993)}]{Spence:1993}%
  \BibitemOpen
  \bibfield  {author} {\bibinfo {author} {\bibfnamefont {J.~R.}\ \bibnamefont
  {Spence}}\ and\ \bibinfo {author} {\bibfnamefont {J.~P.}\ \bibnamefont
  {Vary}},\ }\href {\doibase 10.1103/PhysRevC.47.1282} {\bibfield  {journal}
  {\bibinfo  {journal} {Phys. Rev.}\ }\textbf {\bibinfo {volume} {C47}},\
  \bibinfo {pages} {1282} (\bibinfo {year} {1993})}\BibitemShut {NoStop}%
%%CITATION = PHRVA,C47,1282;%%
\bibitem [{\citenamefont {Crater}\ and\ \citenamefont
  {Van~Alstine}(2004)}]{Crater:2002}%
  \BibitemOpen
  \bibfield  {author} {\bibinfo {author} {\bibfnamefont {H.}~\bibnamefont
  {Crater}}\ and\ \bibinfo {author} {\bibfnamefont {P.}~\bibnamefont
  {Van~Alstine}},\ }\href {\doibase 10.1103/PhysRevD.70.034026} {\bibfield
  {journal} {\bibinfo  {journal} {Phys. Rev.}\ }\textbf {\bibinfo {volume}
  {D70}},\ \bibinfo {pages} {034026} (\bibinfo {year} {2004})},\ \Eprint
  {http://arxiv.org/abs/hep-ph/0208186} {arXiv:hep-ph/0208186 [hep-ph]}
  \BibitemShut {NoStop}%
%%CITATION = HEP-PH/0208186;%%
\bibitem [{\citenamefont {Vary}\ \emph {et~al.}(2010)\citenamefont {Vary},
  \citenamefont {Honkanen}, \citenamefont {Li}, \citenamefont {Maris},
  \citenamefont {Brodsky}, \citenamefont {Harindranath}, \citenamefont
  {de~Teramond}, \citenamefont {Sternberg}, \citenamefont {Ng},\ and\
  \citenamefont {Yang}}]{vary_1stBLFQ}%
  \BibitemOpen
  \bibfield  {author} {\bibinfo {author} {\bibfnamefont {J.~P.}\ \bibnamefont
  {Vary}}, \bibinfo {author} {\bibfnamefont {H.}~\bibnamefont {Honkanen}},
  \bibinfo {author} {\bibfnamefont {J.}~\bibnamefont {Li}}, \bibinfo {author}
  {\bibfnamefont {P.}~\bibnamefont {Maris}}, \bibinfo {author} {\bibfnamefont
  {S.~J.}\ \bibnamefont {Brodsky}}, \bibinfo {author} {\bibfnamefont
  {A.}~\bibnamefont {Harindranath}}, \bibinfo {author} {\bibfnamefont {G.~F.}\
  \bibnamefont {de~Teramond}}, \bibinfo {author} {\bibfnamefont
  {P.}~\bibnamefont {Sternberg}}, \bibinfo {author} {\bibfnamefont {E.~G.}\
  \bibnamefont {Ng}}, \ and\ \bibinfo {author} {\bibfnamefont {C.}~\bibnamefont
  {Yang}},\ }\href {\doibase 10.1103/PhysRevC.81.035205} {\bibfield  {journal}
  {\bibinfo  {journal} {Phys. Rev.}\ }\textbf {\bibinfo {volume} {C81}},\
  \bibinfo {pages} {035205} (\bibinfo {year} {2010})},\ \Eprint
  {http://arxiv.org/abs/0905.1411} {arXiv:0905.1411 [nucl-th]} \BibitemShut
  {NoStop}%
%%CITATION = ARXIV:0905.1411;%%
\bibitem [{\citenamefont {Wiecki}\ \emph {et~al.}(2015)\citenamefont {Wiecki},
  \citenamefont {Li}, \citenamefont {Zhao}, \citenamefont {Maris},\ and\
  \citenamefont {Vary}}]{positronium}%
  \BibitemOpen
  \bibfield  {author} {\bibinfo {author} {\bibfnamefont {P.}~\bibnamefont
  {Wiecki}}, \bibinfo {author} {\bibfnamefont {Y.}~\bibnamefont {Li}}, \bibinfo
  {author} {\bibfnamefont {X.}~\bibnamefont {Zhao}}, \bibinfo {author}
  {\bibfnamefont {P.}~\bibnamefont {Maris}}, \ and\ \bibinfo {author}
  {\bibfnamefont {J.~P.}\ \bibnamefont {Vary}},\ }\href {\doibase
  10.1103/PhysRevD.91.105009} {\bibfield  {journal} {\bibinfo  {journal} {Phys.
  Rev.}\ }\textbf {\bibinfo {volume} {D91}},\ \bibinfo {pages} {105009}
  (\bibinfo {year} {2015})},\ \Eprint {http://arxiv.org/abs/1404.6234}
  {arXiv:1404.6234 [nucl-th]} \BibitemShut {NoStop}%
%%CITATION = ARXIV:1404.6234;%%
\bibitem [{\citenamefont {Li}\ \emph {et~al.}(2016)\citenamefont {Li},
  \citenamefont {Maris}, \citenamefont {Zhao},\ and\ \citenamefont
  {Vary}}]{Li:2015zda}%
  \BibitemOpen
  \bibfield  {author} {\bibinfo {author} {\bibfnamefont {Y.}~\bibnamefont
  {Li}}, \bibinfo {author} {\bibfnamefont {P.}~\bibnamefont {Maris}}, \bibinfo
  {author} {\bibfnamefont {X.}~\bibnamefont {Zhao}}, \ and\ \bibinfo {author}
  {\bibfnamefont {J.~P.}\ \bibnamefont {Vary}},\ }\href {\doibase
  10.1016/j.physletb.2016.04.065} {\bibfield  {journal} {\bibinfo  {journal}
  {Phys. Lett.}\ }\textbf {\bibinfo {volume} {B758}},\ \bibinfo {pages} {118}
  (\bibinfo {year} {2016})},\ \Eprint {http://arxiv.org/abs/1509.07212}
  {arXiv:1509.07212 [hep-ph]} \BibitemShut {NoStop}%
%%CITATION = ARXIV:1509.07212;%%
\bibitem [{\citenamefont {Li}\ \emph {et~al.}(2017)\citenamefont {Li},
  \citenamefont {Maris},\ and\ \citenamefont {Vary}}]{Li:2017mlw}%
  \BibitemOpen
  \bibfield  {author} {\bibinfo {author} {\bibfnamefont {Y.}~\bibnamefont
  {Li}}, \bibinfo {author} {\bibfnamefont {P.}~\bibnamefont {Maris}}, \ and\
  \bibinfo {author} {\bibfnamefont {J.~P.}\ \bibnamefont {Vary}},\ }\href
  {\doibase 10.1103/PhysRevD.96.016022} {\bibfield  {journal} {\bibinfo
  {journal} {Phys. Rev.}\ }\textbf {\bibinfo {volume} {D96}},\ \bibinfo {pages}
  {016022} (\bibinfo {year} {2017})},\ \Eprint
  {http://arxiv.org/abs/1704.06968} {arXiv:1704.06968 [hep-ph]} \BibitemShut
  {NoStop}%
%%CITATION = ARXIV:1704.06968;%%
\bibitem [{\citenamefont {Li}\ \emph {et~al.}(2018{\natexlab{a}})\citenamefont
  {Li}, \citenamefont {Maris},\ and\ \citenamefont {Vary}}]{Yang_frame}%
  \BibitemOpen
  \bibfield  {author} {\bibinfo {author} {\bibfnamefont {Y.}~\bibnamefont
  {Li}}, \bibinfo {author} {\bibfnamefont {P.}~\bibnamefont {Maris}}, \ and\
  \bibinfo {author} {\bibfnamefont {J.~P.}\ \bibnamefont {Vary}},\ }\href
  {\doibase 10.1103/PhysRevD.97.054034} {\bibfield  {journal} {\bibinfo
  {journal} {Phys. Rev.}\ }\textbf {\bibinfo {volume} {D97}},\ \bibinfo {pages}
  {054034} (\bibinfo {year} {2018}{\natexlab{a}})},\ \Eprint
  {http://arxiv.org/abs/1712.03467} {arXiv:1712.03467 [hep-ph]} \BibitemShut
  {NoStop}%
\bibitem [{\citenamefont {Li}\ \emph {et~al.}(2018{\natexlab{b}})\citenamefont
  {Li}, \citenamefont {Li}, \citenamefont {Maris},\ and\ \citenamefont
  {Vary}}]{Li:2018uif}%
  \BibitemOpen
  \bibfield  {author} {\bibinfo {author} {\bibfnamefont {M.}~\bibnamefont
  {Li}}, \bibinfo {author} {\bibfnamefont {Y.}~\bibnamefont {Li}}, \bibinfo
  {author} {\bibfnamefont {P.}~\bibnamefont {Maris}}, \ and\ \bibinfo {author}
  {\bibfnamefont {J.~P.}\ \bibnamefont {Vary}},\ }\href {\doibase
  10.1103/PhysRevD.98.034024} {\bibfield  {journal} {\bibinfo  {journal} {Phys.
  Rev.}\ }\textbf {\bibinfo {volume} {D98}},\ \bibinfo {pages} {034024}
  (\bibinfo {year} {2018}{\natexlab{b}})},\ \Eprint
  {http://arxiv.org/abs/1803.11519} {arXiv:1803.11519 [hep-ph]} \BibitemShut
  {NoStop}%
%%CITATION = ARXIV:1803.11519;%%
\bibitem [{\citenamefont {Adhikari}\ \emph {et~al.}(2019)\citenamefont
  {Adhikari}, \citenamefont {Li}, \citenamefont {Li},\ and\ \citenamefont
  {Vary}}]{Lekha}%
  \BibitemOpen
  \bibfield  {author} {\bibinfo {author} {\bibfnamefont {L.}~\bibnamefont
  {Adhikari}}, \bibinfo {author} {\bibfnamefont {Y.}~\bibnamefont {Li}},
  \bibinfo {author} {\bibfnamefont {M.}~\bibnamefont {Li}}, \ and\ \bibinfo
  {author} {\bibfnamefont {J.~P.}\ \bibnamefont {Vary}},\ }\href {\doibase
  10.1103/PhysRevC.99.035208} {\bibfield  {journal} {\bibinfo  {journal} {Phys.
  Rev.}\ }\textbf {\bibinfo {volume} {C99}},\ \bibinfo {pages} {035208}
  (\bibinfo {year} {2019})},\ \Eprint {http://arxiv.org/abs/1809.06475}
  {arXiv:1809.06475 [hep-ph]} \BibitemShut {NoStop}%
%%CITATION = ARXIV:1809.06475;%%
\bibitem [{\citenamefont {Chen}\ \emph {et~al.}(2017)\citenamefont {Chen},
  \citenamefont {Li}, \citenamefont {Maris}, \citenamefont {Tuchin},\ and\
  \citenamefont {Vary}}]{Chen:VM}%
  \BibitemOpen
  \bibfield  {author} {\bibinfo {author} {\bibfnamefont {G.}~\bibnamefont
  {Chen}}, \bibinfo {author} {\bibfnamefont {Y.}~\bibnamefont {Li}}, \bibinfo
  {author} {\bibfnamefont {P.}~\bibnamefont {Maris}}, \bibinfo {author}
  {\bibfnamefont {K.}~\bibnamefont {Tuchin}}, \ and\ \bibinfo {author}
  {\bibfnamefont {J.~P.}\ \bibnamefont {Vary}},\ }\href {\doibase
  10.1016/j.physletb.2017.04.024} {\bibfield  {journal} {\bibinfo  {journal}
  {Phys. Lett.}\ }\textbf {\bibinfo {volume} {B769}},\ \bibinfo {pages} {477}
  (\bibinfo {year} {2017})},\ \Eprint {http://arxiv.org/abs/1610.04945}
  {arXiv:1610.04945 [nucl-th]} \BibitemShut {NoStop}%
%%CITATION = ARXIV:1610.04945;%%
\bibitem [{\citenamefont {Chen}\ \emph {et~al.}(2019)\citenamefont {Chen},
  \citenamefont {Li}, \citenamefont {Tuchin},\ and\ \citenamefont
  {Vary}}]{Chen:2018vdw}%
  \BibitemOpen
  \bibfield  {author} {\bibinfo {author} {\bibfnamefont {G.}~\bibnamefont
  {Chen}}, \bibinfo {author} {\bibfnamefont {Y.}~\bibnamefont {Li}}, \bibinfo
  {author} {\bibfnamefont {K.}~\bibnamefont {Tuchin}}, \ and\ \bibinfo {author}
  {\bibfnamefont {J.~P.}\ \bibnamefont {Vary}},\ }\href {\doibase
  10.1103/PhysRevC.100.025208} {\bibfield  {journal} {\bibinfo  {journal}
  {Phys. Rev.}\ }\textbf {\bibinfo {volume} {C100}},\ \bibinfo {pages} {025208}
  (\bibinfo {year} {2019})},\ \Eprint {http://arxiv.org/abs/1811.01782}
  {arXiv:1811.01782 [nucl-th]} \BibitemShut {NoStop}%
%%CITATION = ARXIV:1811.01782;%%
\bibitem [{\citenamefont {Tang}\ \emph {et~al.}(2018)\citenamefont {Tang},
  \citenamefont {Li}, \citenamefont {Maris},\ and\ \citenamefont
  {Vary}}]{Tang_bc}%
  \BibitemOpen
  \bibfield  {author} {\bibinfo {author} {\bibfnamefont {S.}~\bibnamefont
  {Tang}}, \bibinfo {author} {\bibfnamefont {Y.}~\bibnamefont {Li}}, \bibinfo
  {author} {\bibfnamefont {P.}~\bibnamefont {Maris}}, \ and\ \bibinfo {author}
  {\bibfnamefont {J.~P.}\ \bibnamefont {Vary}},\ }\href {\doibase
  10.1103/PhysRevD.98.114038} {\bibfield  {journal} {\bibinfo  {journal} {Phys.
  Rev.}\ }\textbf {\bibinfo {volume} {D98}},\ \bibinfo {pages} {114038}
  (\bibinfo {year} {2018})},\ \Eprint {http://arxiv.org/abs/1810.05971}
  {arXiv:1810.05971 [nucl-th]} \BibitemShut {NoStop}%
%%CITATION = ARXIV:1810.05971;%%
\bibitem [{\citenamefont {Jia}\ and\ \citenamefont {Vary}(2019)}]{Jia}%
  \BibitemOpen
  \bibfield  {author} {\bibinfo {author} {\bibfnamefont {S.}~\bibnamefont
  {Jia}}\ and\ \bibinfo {author} {\bibfnamefont {J.~P.}\ \bibnamefont {Vary}},\
  }\href {\doibase 10.1103/PhysRevC.99.035206} {\bibfield  {journal} {\bibinfo
  {journal} {Phys. Rev.}\ }\textbf {\bibinfo {volume} {C99}},\ \bibinfo {pages}
  {035206} (\bibinfo {year} {2019})},\ \Eprint
  {http://arxiv.org/abs/1811.08512} {arXiv:1811.08512 [nucl-th]} \BibitemShut
  {NoStop}%
%%CITATION = ARXIV:1811.08512;%%
\bibitem [{\citenamefont {Lan}\ \emph {et~al.}(2019)\citenamefont {Lan},
  \citenamefont {Mondal}, \citenamefont {Jia}, \citenamefont {Zhao},\ and\
  \citenamefont {Vary}}]{Lan}%
  \BibitemOpen
  \bibfield  {author} {\bibinfo {author} {\bibfnamefont {J.}~\bibnamefont
  {Lan}}, \bibinfo {author} {\bibfnamefont {C.}~\bibnamefont {Mondal}},
  \bibinfo {author} {\bibfnamefont {S.}~\bibnamefont {Jia}}, \bibinfo {author}
  {\bibfnamefont {X.}~\bibnamefont {Zhao}}, \ and\ \bibinfo {author}
  {\bibfnamefont {J.~P.}\ \bibnamefont {Vary}} (\bibinfo {collaboration} {BLFQ
  Collaboration}),\ }\href {\doibase 10.1103/PhysRevLett.122.172001} {\bibfield
   {journal} {\bibinfo  {journal} {Phys. Rev. Lett.}\ }\textbf {\bibinfo
  {volume} {122}},\ \bibinfo {pages} {172001} (\bibinfo {year}
  {2019})}\BibitemShut {NoStop}%
\bibitem [{\citenamefont {Vary}\ \emph {et~al.}(2017)\citenamefont {Vary},
  \citenamefont {Adhikari}, \citenamefont {Chen}, \citenamefont {Li},
  \citenamefont {Li}, \citenamefont {Maris}, \citenamefont {Qian},
  \citenamefont {Spence}, \citenamefont {Tang}, \citenamefont {Tuchin},
  \citenamefont {Yu},\ and\ \citenamefont {Zhao}}]{Vary_2016_proceeding}%
  \BibitemOpen
  \bibfield  {author} {\bibinfo {author} {\bibfnamefont {J.~P.}\ \bibnamefont
  {Vary}}, \bibinfo {author} {\bibfnamefont {L.}~\bibnamefont {Adhikari}},
  \bibinfo {author} {\bibfnamefont {G.}~\bibnamefont {Chen}}, \bibinfo {author}
  {\bibfnamefont {M.}~\bibnamefont {Li}}, \bibinfo {author} {\bibfnamefont
  {Y.}~\bibnamefont {Li}}, \bibinfo {author} {\bibfnamefont {P.}~\bibnamefont
  {Maris}}, \bibinfo {author} {\bibfnamefont {W.}~\bibnamefont {Qian}},
  \bibinfo {author} {\bibfnamefont {J.~R.}\ \bibnamefont {Spence}}, \bibinfo
  {author} {\bibfnamefont {S.}~\bibnamefont {Tang}}, \bibinfo {author}
  {\bibfnamefont {K.}~\bibnamefont {Tuchin}}, \bibinfo {author} {\bibfnamefont
  {A.}~\bibnamefont {Yu}}, \ and\ \bibinfo {author} {\bibfnamefont
  {X.}~\bibnamefont {Zhao}},\ }\bibfield  {booktitle} {\emph {\bibinfo
  {booktitle} {{Proceedings, Theory and Experiment for Hadrons on the
  Light-Front (Light Cone 2016): Lisbon, Portugal, September 5-8, 2016}}},\
  }\href {\doibase 10.1007/s00601-016-1210-1} {\bibfield  {journal} {\bibinfo
  {journal} {Few Body Syst.}\ }\textbf {\bibinfo {volume} {58}},\ \bibinfo
  {pages} {56} (\bibinfo {year} {2017})},\ \Eprint
  {http://arxiv.org/abs/1612.03963} {arXiv:1612.03963 [nucl-th]} \BibitemShut
  {NoStop}%
%%CITATION = ARXIV:1612.03963;%%
\bibitem [{\citenamefont {Vary}\ \emph {et~al.}(2018)\citenamefont {Vary},
  \citenamefont {Adhikari}, \citenamefont {Chen}, \citenamefont {Jia},
  \citenamefont {Li}, \citenamefont {Maris}, \citenamefont {Qian},
  \citenamefont {Spence}, \citenamefont {Tang}, \citenamefont {Tuchin},
  \citenamefont {Yu},\ and\ \citenamefont {Zhao}}]{Vary_2018_proceeding}%
  \BibitemOpen
  \bibfield  {author} {\bibinfo {author} {\bibfnamefont {J.~P.}\ \bibnamefont
  {Vary}}, \bibinfo {author} {\bibfnamefont {L.}~\bibnamefont {Adhikari}},
  \bibinfo {author} {\bibfnamefont {G.}~\bibnamefont {Chen}}, \bibinfo {author}
  {\bibfnamefont {S.}~\bibnamefont {Jia}}, \bibinfo {author} {\bibfnamefont
  {M.}~\bibnamefont {Li}}, \bibinfo {author} {\bibfnamefont {P.}~\bibnamefont
  {Maris}}, \bibinfo {author} {\bibfnamefont {W.}~\bibnamefont {Qian}},
  \bibinfo {author} {\bibfnamefont {J.~R.}\ \bibnamefont {Spence}}, \bibinfo
  {author} {\bibfnamefont {S.}~\bibnamefont {Tang}}, \bibinfo {author}
  {\bibfnamefont {K.}~\bibnamefont {Tuchin}}, \bibinfo {author} {\bibfnamefont
  {A.}~\bibnamefont {Yu}}, \ and\ \bibinfo {author} {\bibfnamefont
  {X.}~\bibnamefont {Zhao}},\ }\bibfield  {booktitle} {\emph {\bibinfo
  {booktitle} {{Proceedings, Frontiers in Light Front Hadron Physics: Theory
  and Experiment (Light Cone 2017): Mumbai, Maharashtra, India, September
  18-22, 2017}}},\ }\href {\doibase 10.1007/s00601-018-1356-0} {\bibfield
  {journal} {\bibinfo  {journal} {Few Body Syst.}\ }\textbf {\bibinfo {volume}
  {59}},\ \bibinfo {pages} {56} (\bibinfo {year} {2018})},\ \Eprint
  {http://arxiv.org/abs/1804.07865} {arXiv:1804.07865 [nucl-th]} \BibitemShut
  {NoStop}%
%%CITATION = ARXIV:1804.07865;%%
\bibitem [{\citenamefont {Klimt}\ \emph {et~al.}(1990)\citenamefont {Klimt},
  \citenamefont {Lutz}, \citenamefont {Vogl},\ and\ \citenamefont
  {Weise}}]{Klimt:1989pm}%
  \BibitemOpen
  \bibfield  {author} {\bibinfo {author} {\bibfnamefont {S.}~\bibnamefont
  {Klimt}}, \bibinfo {author} {\bibfnamefont {M.~F.~M.}\ \bibnamefont {Lutz}},
  \bibinfo {author} {\bibfnamefont {U.}~\bibnamefont {Vogl}}, \ and\ \bibinfo
  {author} {\bibfnamefont {W.}~\bibnamefont {Weise}},\ }\href {\doibase
  10.1016/0375-9474(90)90123-4} {\bibfield  {journal} {\bibinfo  {journal}
  {Nucl. Phys.}\ }\textbf {\bibinfo {volume} {A516}},\ \bibinfo {pages} {429}
  (\bibinfo {year} {1990})}\BibitemShut {NoStop}%
%%CITATION = NUPHA,A516,429;%%
\bibitem [{\citenamefont {Vogl}\ \emph {et~al.}(1990)\citenamefont {Vogl},
  \citenamefont {Lutz}, \citenamefont {Klimt},\ and\ \citenamefont
  {Weise}}]{Vogl:1989ea}%
  \BibitemOpen
  \bibfield  {author} {\bibinfo {author} {\bibfnamefont {U.}~\bibnamefont
  {Vogl}}, \bibinfo {author} {\bibfnamefont {M.~F.~M.}\ \bibnamefont {Lutz}},
  \bibinfo {author} {\bibfnamefont {S.}~\bibnamefont {Klimt}}, \ and\ \bibinfo
  {author} {\bibfnamefont {W.}~\bibnamefont {Weise}},\ }\href {\doibase
  10.1016/0375-9474(90)90124-5} {\bibfield  {journal} {\bibinfo  {journal}
  {Nucl. Phys.}\ }\textbf {\bibinfo {volume} {A516}},\ \bibinfo {pages} {469}
  (\bibinfo {year} {1990})}\BibitemShut {NoStop}%
%%CITATION = NUPHA,A516,469;%%
\bibitem [{\citenamefont {Vogl}\ and\ \citenamefont
  {Weise}(1991)}]{Vogl:1991qt}%
  \BibitemOpen
  \bibfield  {author} {\bibinfo {author} {\bibfnamefont {U.}~\bibnamefont
  {Vogl}}\ and\ \bibinfo {author} {\bibfnamefont {W.}~\bibnamefont {Weise}},\
  }\href {\doibase 10.1016/0146-6410(91)90005-9} {\bibfield  {journal}
  {\bibinfo  {journal} {Prog. Part. Nucl. Phys.}\ }\textbf {\bibinfo {volume}
  {27}},\ \bibinfo {pages} {195} (\bibinfo {year} {1991})}\BibitemShut
  {NoStop}%
%%CITATION = PPNPD,27,195;%%
\bibitem [{\citenamefont {de~Teramond}\ and\ \citenamefont
  {Brodsky}(2009)}]{deTeramond:2008ht}%
  \BibitemOpen
  \bibfield  {author} {\bibinfo {author} {\bibfnamefont {G.~F.}\ \bibnamefont
  {de~Teramond}}\ and\ \bibinfo {author} {\bibfnamefont {S.~J.}\ \bibnamefont
  {Brodsky}},\ }\href {\doibase 10.1103/PhysRevLett.102.081601} {\bibfield
  {journal} {\bibinfo  {journal} {Phys. Rev. Lett.}\ }\textbf {\bibinfo
  {volume} {102}},\ \bibinfo {pages} {081601} (\bibinfo {year} {2009})},\
  \Eprint {http://arxiv.org/abs/0809.4899} {arXiv:0809.4899 [hep-ph]}
  \BibitemShut {NoStop}%
%%CITATION = ARXIV:0809.4899;%%
\bibitem [{\citenamefont {Tanabashi}\ \emph {et~al.}(2018)\citenamefont
  {Tanabashi}, \citenamefont {Hagiwara}, \citenamefont {Hikasa}, \citenamefont
  {Nakamura}, \citenamefont {Sumino}, \citenamefont {Takahashi}, \citenamefont
  {Tanaka}, \citenamefont {Agashe}, \citenamefont {Aielli}, \citenamefont
  {Amsler} \emph {et~al.}}]{PhysRevD.98.030001}%
  \BibitemOpen
  \bibfield  {author} {\bibinfo {author} {\bibfnamefont {M.}~\bibnamefont
  {Tanabashi}}, \bibinfo {author} {\bibfnamefont {K.}~\bibnamefont {Hagiwara}},
  \bibinfo {author} {\bibfnamefont {K.}~\bibnamefont {Hikasa}}, \bibinfo
  {author} {\bibfnamefont {K.}~\bibnamefont {Nakamura}}, \bibinfo {author}
  {\bibfnamefont {Y.}~\bibnamefont {Sumino}}, \bibinfo {author} {\bibfnamefont
  {F.}~\bibnamefont {Takahashi}}, \bibinfo {author} {\bibfnamefont
  {J.}~\bibnamefont {Tanaka}}, \bibinfo {author} {\bibfnamefont
  {K.}~\bibnamefont {Agashe}}, \bibinfo {author} {\bibfnamefont
  {G.}~\bibnamefont {Aielli}}, \bibinfo {author} {\bibfnamefont
  {C.}~\bibnamefont {Amsler}},  \emph {et~al.} (\bibinfo {collaboration}
  {Particle Data Group}),\ }\href {\doibase 10.1103/PhysRevD.98.030001}
  {\bibfield  {journal} {\bibinfo  {journal} {Phys. Rev. D}\ }\textbf {\bibinfo
  {volume} {98}},\ \bibinfo {pages} {030001} (\bibinfo {year}
  {2018})}\BibitemShut {NoStop}%
\bibitem [{\citenamefont {Li}(2019)}]{ML_thesis}%
  \BibitemOpen
  \bibfield  {author} {\bibinfo {author} {\bibfnamefont {M.}~\bibnamefont
  {Li}},\ }\emph {\bibinfo {title} {{Non-perturbative applications of quantum
  chromodynamics}}},\ \href@noop {} {Ph.D. thesis},\ \bibinfo  {school} {Iowa
  State University} (\bibinfo {year} {2019})\BibitemShut {NoStop}%
\bibitem [{\citenamefont {Jia}()}]{Jia_revised}%
  \BibitemOpen
  \bibfield  {author} {\bibinfo {author} {\bibfnamefont {S.}~\bibnamefont
  {Jia}},\ }\href@noop {} {}\bibinfo {note} {In preparation}\BibitemShut
  {NoStop}%
\bibitem [{\citenamefont {Drell}\ and\ \citenamefont {Yan}(1970)}]{DrellYan}%
  \BibitemOpen
  \bibfield  {author} {\bibinfo {author} {\bibfnamefont {S.~D.}\ \bibnamefont
  {Drell}}\ and\ \bibinfo {author} {\bibfnamefont {T.-M.}\ \bibnamefont
  {Yan}},\ }\href {\doibase 10.1103/PhysRevLett.24.181} {\bibfield  {journal}
  {\bibinfo  {journal} {Phys. Rev. Lett.}\ }\textbf {\bibinfo {volume} {24}},\
  \bibinfo {pages} {181} (\bibinfo {year} {1970})}\BibitemShut {NoStop}%
%%CITATION = PRLTA,24,181;%%
\bibitem [{\citenamefont {Grach}\ and\ \citenamefont
  {Kondratyuk}(1984)}]{GK_1984}%
  \BibitemOpen
  \bibfield  {author} {\bibinfo {author} {\bibfnamefont {I.~L.}\ \bibnamefont
  {Grach}}\ and\ \bibinfo {author} {\bibfnamefont {L.~A.}\ \bibnamefont
  {Kondratyuk}},\ }\href@noop {} {\bibfield  {journal} {\bibinfo  {journal}
  {Sov. J. Nucl. Phys.}\ }\textbf {\bibinfo {volume} {39}},\ \bibinfo {pages}
  {198} (\bibinfo {year} {1984})},\ \bibinfo {note} {[Yad.
  Fiz.39,316(1984)]}\BibitemShut {NoStop}%
%%CITATION = SJNCA,39,198;%%
\bibitem [{\citenamefont {Cardarelli}\ \emph {et~al.}(1995)\citenamefont
  {Cardarelli}, \citenamefont {Grach}, \citenamefont {Narodetsky},
  \citenamefont {Salme},\ and\ \citenamefont {Simula}}]{GK_1995}%
  \BibitemOpen
  \bibfield  {author} {\bibinfo {author} {\bibfnamefont {F.}~\bibnamefont
  {Cardarelli}}, \bibinfo {author} {\bibfnamefont {I.~L.}\ \bibnamefont
  {Grach}}, \bibinfo {author} {\bibfnamefont {I.~M.}\ \bibnamefont
  {Narodetsky}}, \bibinfo {author} {\bibfnamefont {G.}~\bibnamefont {Salme}}, \
  and\ \bibinfo {author} {\bibfnamefont {S.}~\bibnamefont {Simula}},\ }\href
  {\doibase 10.1016/0370-2693(95)00230-I} {\bibfield  {journal} {\bibinfo
  {journal} {Phys. Lett.}\ }\textbf {\bibinfo {volume} {B349}},\ \bibinfo
  {pages} {393} (\bibinfo {year} {1995})},\ \Eprint
  {http://arxiv.org/abs/hep-ph/9502360} {arXiv:hep-ph/9502360 [hep-ph]}
  \BibitemShut {NoStop}%
%%CITATION = HEP-PH/9502360;%%
\bibitem [{\citenamefont {Volmer}\ \emph {et~al.}(2001)\citenamefont {Volmer},
  \citenamefont {Abbott}, \citenamefont {Anklin}, \citenamefont {Armstrong},
  \citenamefont {Arrington}, \citenamefont {Assamagan}, \citenamefont {Avery},
  \citenamefont {Baker}, \citenamefont {Blok}, \citenamefont {Bochna} \emph
  {et~al.}}]{Volmer}%
  \BibitemOpen
  \bibfield  {author} {\bibinfo {author} {\bibfnamefont {J.}~\bibnamefont
  {Volmer}}, \bibinfo {author} {\bibfnamefont {D.}~\bibnamefont {Abbott}},
  \bibinfo {author} {\bibfnamefont {H.}~\bibnamefont {Anklin}}, \bibinfo
  {author} {\bibfnamefont {C.}~\bibnamefont {Armstrong}}, \bibinfo {author}
  {\bibfnamefont {J.}~\bibnamefont {Arrington}}, \bibinfo {author}
  {\bibfnamefont {K.}~\bibnamefont {Assamagan}}, \bibinfo {author}
  {\bibfnamefont {S.}~\bibnamefont {Avery}}, \bibinfo {author} {\bibfnamefont
  {O.~K.}\ \bibnamefont {Baker}}, \bibinfo {author} {\bibfnamefont {H.~P.}\
  \bibnamefont {Blok}}, \bibinfo {author} {\bibfnamefont {C.}~\bibnamefont
  {Bochna}},  \emph {et~al.} (\bibinfo {collaboration} {Jefferson Lab F(pi)}),\
  }\href {\doibase 10.1103/PhysRevLett.86.1713} {\bibfield  {journal} {\bibinfo
   {journal} {Phys. Rev. Lett.}\ }\textbf {\bibinfo {volume} {86}},\ \bibinfo
  {pages} {1713} (\bibinfo {year} {2001})},\ \Eprint
  {http://arxiv.org/abs/nucl-ex/0010009} {arXiv:nucl-ex/0010009 [nucl-ex]}
  \BibitemShut {NoStop}%
%%CITATION = NUCL-EX/0010009;%%
\bibitem [{\citenamefont {Amendolia}\ \emph {et~al.}(1984)\citenamefont
  {Amendolia}, \citenamefont {Badelek}, \citenamefont {Batignani},
  \citenamefont {Beck}, \citenamefont {Bedeschi}, \citenamefont {Bellamy},
  \citenamefont {Bertolucci}, \citenamefont {Bettonic}, \citenamefont
  {Bilokon}, \citenamefont {Bologna} \emph {et~al.}}]{Amendolia_1984}%
  \BibitemOpen
  \bibfield  {author} {\bibinfo {author} {\bibfnamefont {S.~R.}\ \bibnamefont
  {Amendolia}}, \bibinfo {author} {\bibfnamefont {B.}~\bibnamefont {Badelek}},
  \bibinfo {author} {\bibfnamefont {G.}~\bibnamefont {Batignani}}, \bibinfo
  {author} {\bibfnamefont {G.~A.}\ \bibnamefont {Beck}}, \bibinfo {author}
  {\bibfnamefont {F.}~\bibnamefont {Bedeschi}}, \bibinfo {author}
  {\bibfnamefont {E.~H.}\ \bibnamefont {Bellamy}}, \bibinfo {author}
  {\bibfnamefont {E.}~\bibnamefont {Bertolucci}}, \bibinfo {author}
  {\bibfnamefont {D.}~\bibnamefont {Bettonic}}, \bibinfo {author}
  {\bibfnamefont {H.}~\bibnamefont {Bilokon}}, \bibinfo {author} {\bibfnamefont
  {G.}~\bibnamefont {Bologna}},  \emph {et~al.},\ }\href {\doibase
  10.1016/0370-2693(84)90655-5} {\bibfield  {journal} {\bibinfo  {journal}
  {Phys. Lett.}\ }\textbf {\bibinfo {volume} {146B}},\ \bibinfo {pages} {116}
  (\bibinfo {year} {1984})}\BibitemShut {NoStop}%
%%CITATION = PHLTA,146B,116;%%
\bibitem [{\citenamefont {Amendolia}\ \emph {et~al.}(1986)\citenamefont
  {Amendolia}, \citenamefont {Arik}, \citenamefont {Badelek}, \citenamefont
  {Batignani}, \citenamefont {Beck}, \citenamefont {Bedeschi}, \citenamefont
  {Bellamy}, \citenamefont {Bertolucci}, \citenamefont {Bettonic},
  \citenamefont {Bilokon} \emph {et~al.}}]{Amendolia_1986}%
  \BibitemOpen
  \bibfield  {author} {\bibinfo {author} {\bibfnamefont {S.~R.}\ \bibnamefont
  {Amendolia}}, \bibinfo {author} {\bibfnamefont {M.}~\bibnamefont {Arik}},
  \bibinfo {author} {\bibfnamefont {B.}~\bibnamefont {Badelek}}, \bibinfo
  {author} {\bibfnamefont {G.}~\bibnamefont {Batignani}}, \bibinfo {author}
  {\bibfnamefont {G.~A.}\ \bibnamefont {Beck}}, \bibinfo {author}
  {\bibfnamefont {F.}~\bibnamefont {Bedeschi}}, \bibinfo {author}
  {\bibfnamefont {E.~H.}\ \bibnamefont {Bellamy}}, \bibinfo {author}
  {\bibfnamefont {E.}~\bibnamefont {Bertolucci}}, \bibinfo {author}
  {\bibfnamefont {D.}~\bibnamefont {Bettonic}}, \bibinfo {author}
  {\bibfnamefont {H.}~\bibnamefont {Bilokon}},  \emph {et~al.} (\bibinfo
  {collaboration} {NA7}),\ }\bibfield  {booktitle} {\emph {\bibinfo {booktitle}
  {{Proceedings, 23RD International Conference on High Energy Physics, JULY
  16-23, 1986, Berkeley, CA}}},\ }\href {\doibase 10.1016/0550-3213(86)90437-2}
  {\bibfield  {journal} {\bibinfo  {journal} {Nucl. Phys.}\ }\textbf {\bibinfo
  {volume} {B277}},\ \bibinfo {pages} {168} (\bibinfo {year}
  {1986})}\BibitemShut {NoStop}%
%%CITATION = NUPHA,B277,168;%%
\bibitem [{\citenamefont {Owen}\ \emph {et~al.}(2015)\citenamefont {Owen},
  \citenamefont {Kamleh}, \citenamefont {Leinweber}, \citenamefont {Menadue},\
  and\ \citenamefont {Mahbub}}]{Owen:2015}%
  \BibitemOpen
  \bibfield  {author} {\bibinfo {author} {\bibfnamefont {B.}~\bibnamefont
  {Owen}}, \bibinfo {author} {\bibfnamefont {W.}~\bibnamefont {Kamleh}},
  \bibinfo {author} {\bibfnamefont {D.}~\bibnamefont {Leinweber}}, \bibinfo
  {author} {\bibfnamefont {B.}~\bibnamefont {Menadue}}, \ and\ \bibinfo
  {author} {\bibfnamefont {S.}~\bibnamefont {Mahbub}},\ }\href {\doibase
  10.1103/PhysRevD.91.074503} {\bibfield  {journal} {\bibinfo  {journal} {Phys.
  Rev.}\ }\textbf {\bibinfo {volume} {D91}},\ \bibinfo {pages} {074503}
  (\bibinfo {year} {2015})},\ \Eprint {http://arxiv.org/abs/1501.02561}
  {arXiv:1501.02561 [hep-lat]} \BibitemShut {NoStop}%
%%CITATION = ARXIV:1501.02561;%%
\bibitem [{\citenamefont {Bhagwat}\ and\ \citenamefont
  {Maris}(2008)}]{Bhagwat_Maris:2008}%
  \BibitemOpen
  \bibfield  {author} {\bibinfo {author} {\bibfnamefont {M.~S.}\ \bibnamefont
  {Bhagwat}}\ and\ \bibinfo {author} {\bibfnamefont {P.}~\bibnamefont
  {Maris}},\ }\href {\doibase 10.1103/PhysRevC.77.025203} {\bibfield  {journal}
  {\bibinfo  {journal} {Phys. Rev.}\ }\textbf {\bibinfo {volume} {C77}},\
  \bibinfo {pages} {025203} (\bibinfo {year} {2008})},\ \Eprint
  {http://arxiv.org/abs/nucl-th/0612069} {arXiv:nucl-th/0612069 [nucl-th]}
  \BibitemShut {NoStop}%
%%CITATION = NUCL-TH/0612069;%%
\bibitem [{\citenamefont {Choi}\ and\ \citenamefont {Ji}(2004)}]{Choi:rho}%
  \BibitemOpen
  \bibfield  {author} {\bibinfo {author} {\bibfnamefont {H.-M.}\ \bibnamefont
  {Choi}}\ and\ \bibinfo {author} {\bibfnamefont {C.-R.}\ \bibnamefont {Ji}},\
  }\href {\doibase 10.1103/PhysRevD.70.053015} {\bibfield  {journal} {\bibinfo
  {journal} {Phys. Rev.}\ }\textbf {\bibinfo {volume} {D70}},\ \bibinfo {pages}
  {053015} (\bibinfo {year} {2004})},\ \Eprint
  {http://arxiv.org/abs/hep-ph/0402114} {arXiv:hep-ph/0402114 [hep-ph]}
  \BibitemShut {NoStop}%
%%CITATION = HEP-PH/0402114;%%
\bibitem [{\citenamefont {Carrillo-Serrano}\ \emph {et~al.}(2015)\citenamefont
  {Carrillo-Serrano}, \citenamefont {Bentz}, \citenamefont {Cloët},\ and\
  \citenamefont {Thomas}}]{Carrillo_NJL:2015}%
  \BibitemOpen
  \bibfield  {author} {\bibinfo {author} {\bibfnamefont {M.~E.}\ \bibnamefont
  {Carrillo-Serrano}}, \bibinfo {author} {\bibfnamefont {W.}~\bibnamefont
  {Bentz}}, \bibinfo {author} {\bibfnamefont {I.~C.}\ \bibnamefont {Cloët}}, \
  and\ \bibinfo {author} {\bibfnamefont {A.~W.}\ \bibnamefont {Thomas}},\
  }\href {\doibase 10.1103/PhysRevC.92.015212} {\bibfield  {journal} {\bibinfo
  {journal} {Phys. Rev.}\ }\textbf {\bibinfo {volume} {C92}},\ \bibinfo {pages}
  {015212} (\bibinfo {year} {2015})},\ \Eprint
  {http://arxiv.org/abs/1504.08119} {arXiv:1504.08119 [nucl-th]} \BibitemShut
  {NoStop}%
%%CITATION = ARXIV:1504.08119;%%
\bibitem [{\citenamefont {Ninomiya}\ \emph {et~al.}(2015)\citenamefont
  {Ninomiya}, \citenamefont {Bentz},\ and\ \citenamefont
  {Cloët}}]{Ninomiya_NJL:2015}%
  \BibitemOpen
  \bibfield  {author} {\bibinfo {author} {\bibfnamefont {Y.}~\bibnamefont
  {Ninomiya}}, \bibinfo {author} {\bibfnamefont {W.}~\bibnamefont {Bentz}}, \
  and\ \bibinfo {author} {\bibfnamefont {I.~C.}\ \bibnamefont {Cloët}},\
  }\href {\doibase 10.1103/PhysRevC.91.025202} {\bibfield  {journal} {\bibinfo
  {journal} {Phys. Rev.}\ }\textbf {\bibinfo {volume} {C91}},\ \bibinfo {pages}
  {025202} (\bibinfo {year} {2015})},\ \Eprint {http://arxiv.org/abs/1406.7212}
  {arXiv:1406.7212 [nucl-th]} \BibitemShut {NoStop}%
%%CITATION = ARXIV:1406.7212;%%
\bibitem [{\citenamefont {Dokshitzer}(1977)}]{Dokshitzer:1977sg}%
  \BibitemOpen
  \bibfield  {author} {\bibinfo {author} {\bibfnamefont {Y.~L.}\ \bibnamefont
  {Dokshitzer}},\ }\href@noop {} {\bibfield  {journal} {\bibinfo  {journal}
  {Sov. Phys. JETP}\ }\textbf {\bibinfo {volume} {46}},\ \bibinfo {pages} {641}
  (\bibinfo {year} {1977})},\ \bibinfo {note} {[Zh. Eksp. Teor.
  Fiz.73,1216(1977)]}\BibitemShut {NoStop}%
%%CITATION = SPHJA,46,641;%%
\bibitem [{\citenamefont {Gribov}\ and\ \citenamefont
  {Lipatov}(1972)}]{Gribov:1972ri}%
  \BibitemOpen
  \bibfield  {author} {\bibinfo {author} {\bibfnamefont {V.~N.}\ \bibnamefont
  {Gribov}}\ and\ \bibinfo {author} {\bibfnamefont {L.~N.}\ \bibnamefont
  {Lipatov}},\ }\href@noop {} {\bibfield  {journal} {\bibinfo  {journal} {Sov.
  J. Nucl. Phys.}\ }\textbf {\bibinfo {volume} {15}},\ \bibinfo {pages} {438}
  (\bibinfo {year} {1972})},\ \bibinfo {note} {[Yad.
  Fiz.15,781(1972)]}\BibitemShut {NoStop}%
%%CITATION = SJNCA,15,438;%%
\bibitem [{\citenamefont {Altarelli}\ and\ \citenamefont
  {Parisi}(1977)}]{Altarelli:1977zs}%
  \BibitemOpen
  \bibfield  {author} {\bibinfo {author} {\bibfnamefont {G.}~\bibnamefont
  {Altarelli}}\ and\ \bibinfo {author} {\bibfnamefont {G.}~\bibnamefont
  {Parisi}},\ }\href {\doibase 10.1016/0550-3213(77)90384-4} {\bibfield
  {journal} {\bibinfo  {journal} {Nucl. Phys.}\ }\textbf {\bibinfo {volume}
  {B126}},\ \bibinfo {pages} {298} (\bibinfo {year} {1977})}\BibitemShut
  {NoStop}%
%%CITATION = NUPHA,B126,298;%%
\bibitem [{\citenamefont {Salam}\ and\ \citenamefont
  {Rojo}(2009)}]{Salam:2008qg}%
  \BibitemOpen
  \bibfield  {author} {\bibinfo {author} {\bibfnamefont {G.~P.}\ \bibnamefont
  {Salam}}\ and\ \bibinfo {author} {\bibfnamefont {J.}~\bibnamefont {Rojo}},\
  }\href {\doibase 10.1016/j.cpc.2008.08.010} {\bibfield  {journal} {\bibinfo
  {journal} {Comput. Phys. Commun.}\ }\textbf {\bibinfo {volume} {180}},\
  \bibinfo {pages} {120} (\bibinfo {year} {2009})},\ \Eprint
  {http://arxiv.org/abs/0804.3755} {arXiv:0804.3755 [hep-ph]} \BibitemShut
  {NoStop}%
%%CITATION = ARXIV:0804.3755;%%
\bibitem [{\citenamefont {Conway}\ \emph {et~al.}(1989)\citenamefont {Conway},
  \citenamefont {Adolphsen}, \citenamefont {Alexander}, \citenamefont
  {Anderson}, \citenamefont {Heinrich}, \citenamefont {Pilcher}, \citenamefont
  {Possoz}, \citenamefont {Rosenberg}, \citenamefont {Biino}, \citenamefont
  {Greenhalgh} \emph {et~al.}}]{E615_Conway}%
  \BibitemOpen
  \bibfield  {author} {\bibinfo {author} {\bibfnamefont {J.~S.}\ \bibnamefont
  {Conway}}, \bibinfo {author} {\bibfnamefont {C.~E.}\ \bibnamefont
  {Adolphsen}}, \bibinfo {author} {\bibfnamefont {J.~P.}\ \bibnamefont
  {Alexander}}, \bibinfo {author} {\bibfnamefont {K.~J.}\ \bibnamefont
  {Anderson}}, \bibinfo {author} {\bibfnamefont {J.~G.}\ \bibnamefont
  {Heinrich}}, \bibinfo {author} {\bibfnamefont {J.~E.}\ \bibnamefont
  {Pilcher}}, \bibinfo {author} {\bibfnamefont {A.}~\bibnamefont {Possoz}},
  \bibinfo {author} {\bibfnamefont {E.~I.}\ \bibnamefont {Rosenberg}}, \bibinfo
  {author} {\bibfnamefont {C.}~\bibnamefont {Biino}}, \bibinfo {author}
  {\bibfnamefont {J.~F.}\ \bibnamefont {Greenhalgh}},  \emph {et~al.},\ }\href
  {\doibase 10.1103/PhysRevD.39.92} {\bibfield  {journal} {\bibinfo  {journal}
  {Phys. Rev.}\ }\textbf {\bibinfo {volume} {D39}},\ \bibinfo {pages} {92}
  (\bibinfo {year} {1989})}\BibitemShut {NoStop}%
%%CITATION = PHRVA,D39,92;%%
\bibitem [{\citenamefont {Lepage}\ and\ \citenamefont
  {Brodsky}(1980)}]{Lepage:1980}%
  \BibitemOpen
  \bibfield  {author} {\bibinfo {author} {\bibfnamefont {G.~P.}\ \bibnamefont
  {Lepage}}\ and\ \bibinfo {author} {\bibfnamefont {S.~J.}\ \bibnamefont
  {Brodsky}},\ }\href {\doibase 10.1103/PhysRevD.22.2157} {\bibfield  {journal}
  {\bibinfo  {journal} {Phys. Rev.}\ }\textbf {\bibinfo {volume} {D22}},\
  \bibinfo {pages} {2157} (\bibinfo {year} {1980})}\BibitemShut {NoStop}%
%%CITATION = PHRVA,D22,2157;%%
\bibitem [{\citenamefont {Efremov}\ and\ \citenamefont
  {Radyushkin}(1980{\natexlab{a}})}]{Efremov:1980}%
  \BibitemOpen
  \bibfield  {author} {\bibinfo {author} {\bibfnamefont {A.~V.}\ \bibnamefont
  {Efremov}}\ and\ \bibinfo {author} {\bibfnamefont {A.~V.}\ \bibnamefont
  {Radyushkin}},\ }\href {\doibase 10.1007/BF01032111} {\bibfield  {journal}
  {\bibinfo  {journal} {Theor. Math. Phys.}\ }\textbf {\bibinfo {volume}
  {42}},\ \bibinfo {pages} {97} (\bibinfo {year} {1980}{\natexlab{a}})},\
  \bibinfo {note} {[Teor. Mat. Fiz.42,147(1980)]}\BibitemShut {NoStop}%
%%CITATION = TMPHA,42,97;%%
\bibitem [{\citenamefont {Efremov}\ and\ \citenamefont
  {Radyushkin}(1980{\natexlab{b}})}]{Efremov:1980_v2}%
  \BibitemOpen
  \bibfield  {author} {\bibinfo {author} {\bibfnamefont {A.~V.}\ \bibnamefont
  {Efremov}}\ and\ \bibinfo {author} {\bibfnamefont {A.~V.}\ \bibnamefont
  {Radyushkin}},\ }\href {\doibase 10.1016/0370-2693(80)90869-2} {\bibfield
  {journal} {\bibinfo  {journal} {Phys. Lett.}\ }\textbf {\bibinfo {volume}
  {94B}},\ \bibinfo {pages} {245} (\bibinfo {year}
  {1980}{\natexlab{b}})}\BibitemShut {NoStop}%
%%CITATION = PHLTA,94B,245;%%
\bibitem [{\citenamefont {Aitala}\ \emph {et~al.}(2001)\citenamefont {Aitala},
  \citenamefont {Amato}, \citenamefont {Anjos}, \citenamefont {Appel},
  \citenamefont {Ashery}, \citenamefont {Banerjee}, \citenamefont {Bediaga},
  \citenamefont {Blaylock}, \citenamefont {Bracker}, \citenamefont {Burchat}
  \emph {et~al.}}]{E791}%
  \BibitemOpen
  \bibfield  {author} {\bibinfo {author} {\bibfnamefont {E.~M.}\ \bibnamefont
  {Aitala}}, \bibinfo {author} {\bibfnamefont {S.}~\bibnamefont {Amato}},
  \bibinfo {author} {\bibfnamefont {J.~C.}\ \bibnamefont {Anjos}}, \bibinfo
  {author} {\bibfnamefont {J.~A.}\ \bibnamefont {Appel}}, \bibinfo {author}
  {\bibfnamefont {D.}~\bibnamefont {Ashery}}, \bibinfo {author} {\bibfnamefont
  {S.}~\bibnamefont {Banerjee}}, \bibinfo {author} {\bibfnamefont
  {I.}~\bibnamefont {Bediaga}}, \bibinfo {author} {\bibfnamefont
  {G.}~\bibnamefont {Blaylock}}, \bibinfo {author} {\bibfnamefont {S.~B.}\
  \bibnamefont {Bracker}}, \bibinfo {author} {\bibfnamefont {P.~R.}\
  \bibnamefont {Burchat}},  \emph {et~al.} (\bibinfo {collaboration} {E791}),\
  }\href {\doibase 10.1103/PhysRevLett.86.4768} {\bibfield  {journal} {\bibinfo
   {journal} {Phys. Rev. Lett.}\ }\textbf {\bibinfo {volume} {86}},\ \bibinfo
  {pages} {4768} (\bibinfo {year} {2001})},\ \Eprint
  {http://arxiv.org/abs/hep-ex/0010043} {arXiv:hep-ex/0010043 [hep-ex]}
  \BibitemShut {NoStop}%
%%CITATION = HEP-EX/0010043;%%
\bibitem [{\citenamefont {Ruiz~Arriola}\ and\ \citenamefont
  {Broniowski}(2002)}]{Ruiz}%
  \BibitemOpen
  \bibfield  {author} {\bibinfo {author} {\bibfnamefont {E.}~\bibnamefont
  {Ruiz~Arriola}}\ and\ \bibinfo {author} {\bibfnamefont {W.}~\bibnamefont
  {Broniowski}},\ }\href {\doibase 10.1103/PhysRevD.66.094016} {\bibfield
  {journal} {\bibinfo  {journal} {Phys. Rev.}\ }\textbf {\bibinfo {volume}
  {D66}},\ \bibinfo {pages} {094016} (\bibinfo {year} {2002})},\ \Eprint
  {http://arxiv.org/abs/hep-ph/0207266} {arXiv:hep-ph/0207266 [hep-ph]}
  \BibitemShut {NoStop}%
%%CITATION = HEP-PH/0207266;%%
\bibitem [{\citenamefont {Brodsky}\ and\ \citenamefont
  {de~Teramond}(2008)}]{holography_pion}%
  \BibitemOpen
  \bibfield  {author} {\bibinfo {author} {\bibfnamefont {S.~J.}\ \bibnamefont
  {Brodsky}}\ and\ \bibinfo {author} {\bibfnamefont {G.~F.}\ \bibnamefont
  {de~Teramond}},\ }\href {\doibase 10.1103/PhysRevD.77.056007} {\bibfield
  {journal} {\bibinfo  {journal} {Phys. Rev.}\ }\textbf {\bibinfo {volume}
  {D77}},\ \bibinfo {pages} {056007} (\bibinfo {year} {2008})},\ \Eprint
  {http://arxiv.org/abs/0707.3859} {arXiv:0707.3859 [hep-ph]} \BibitemShut
  {NoStop}%
%%CITATION = ARXIV:0707.3859;%%
\bibitem [{\citenamefont {Stefanis}\ and\ \citenamefont
  {Pimikov}(2016)}]{Stefanis:2015qha}%
  \BibitemOpen
  \bibfield  {author} {\bibinfo {author} {\bibfnamefont {N.}~\bibnamefont
  {Stefanis}}\ and\ \bibinfo {author} {\bibfnamefont {A.}~\bibnamefont
  {Pimikov}},\ }\href {\doibase 10.1016/j.nuclphysa.2015.11.002} {\bibfield
  {journal} {\bibinfo  {journal} {Nucl. Phys. A}\ }\textbf {\bibinfo {volume}
  {945}},\ \bibinfo {pages} {248} (\bibinfo {year} {2016})},\ \Eprint
  {http://arxiv.org/abs/1506.01302} {arXiv:1506.01302 [hep-ph]} \BibitemShut
  {NoStop}%
\bibitem [{\citenamefont {Ball}\ and\ \citenamefont {Zwicky}(2005)}]{Ball_sr}%
  \BibitemOpen
  \bibfield  {author} {\bibinfo {author} {\bibfnamefont {P.}~\bibnamefont
  {Ball}}\ and\ \bibinfo {author} {\bibfnamefont {R.}~\bibnamefont {Zwicky}},\
  }\href {\doibase 10.1103/PhysRevD.71.014015} {\bibfield  {journal} {\bibinfo
  {journal} {Phys. Rev.}\ }\textbf {\bibinfo {volume} {D71}},\ \bibinfo {pages}
  {014015} (\bibinfo {year} {2005})},\ \Eprint
  {http://arxiv.org/abs/hep-ph/0406232} {arXiv:hep-ph/0406232 [hep-ph]}
  \BibitemShut {NoStop}%
%%CITATION = HEP-PH/0406232;%%
\bibitem [{\citenamefont {Forshaw}\ and\ \citenamefont
  {Sandapen}(2010)}]{Forshaw_rho_meson}%
  \BibitemOpen
  \bibfield  {author} {\bibinfo {author} {\bibfnamefont {J.~R.}\ \bibnamefont
  {Forshaw}}\ and\ \bibinfo {author} {\bibfnamefont {R.}~\bibnamefont
  {Sandapen}},\ }\href {\doibase 10.1007/JHEP11(2010)037} {\bibfield  {journal}
  {\bibinfo  {journal} {JHEP}\ }\textbf {\bibinfo {volume} {11}},\ \bibinfo
  {pages} {037} (\bibinfo {year} {2010})},\ \Eprint
  {http://arxiv.org/abs/1007.1990} {arXiv:1007.1990 [hep-ph]} \BibitemShut
  {NoStop}%
%%CITATION = ARXIV:1007.1990;%%
\bibitem [{\citenamefont {Gao}\ \emph {et~al.}(2014)\citenamefont {Gao},
  \citenamefont {Chang}, \citenamefont {Liu}, \citenamefont {Roberts},\ and\
  \citenamefont {Schmidt}}]{Gao_bse}%
  \BibitemOpen
  \bibfield  {author} {\bibinfo {author} {\bibfnamefont {F.}~\bibnamefont
  {Gao}}, \bibinfo {author} {\bibfnamefont {L.}~\bibnamefont {Chang}}, \bibinfo
  {author} {\bibfnamefont {Y.-X.}\ \bibnamefont {Liu}}, \bibinfo {author}
  {\bibfnamefont {C.~D.}\ \bibnamefont {Roberts}}, \ and\ \bibinfo {author}
  {\bibfnamefont {S.~M.}\ \bibnamefont {Schmidt}},\ }\href {\doibase
  10.1103/PhysRevD.90.014011} {\bibfield  {journal} {\bibinfo  {journal} {Phys.
  Rev.}\ }\textbf {\bibinfo {volume} {D90}},\ \bibinfo {pages} {014011}
  (\bibinfo {year} {2014})},\ \Eprint {http://arxiv.org/abs/1405.0289}
  {arXiv:1405.0289 [nucl-th]} \BibitemShut {NoStop}%
%%CITATION = ARXIV:1405.0289;%%
\bibitem [{\citenamefont {Bakulev}\ and\ \citenamefont
  {Mikhailov}(1998)}]{Bakulev_rho}%
  \BibitemOpen
  \bibfield  {author} {\bibinfo {author} {\bibfnamefont {A.~P.}\ \bibnamefont
  {Bakulev}}\ and\ \bibinfo {author} {\bibfnamefont {S.~V.}\ \bibnamefont
  {Mikhailov}},\ }\href {\doibase 10.1016/S0370-2693(98)00868-5} {\bibfield
  {journal} {\bibinfo  {journal} {Phys. Lett.}\ }\textbf {\bibinfo {volume}
  {B436}},\ \bibinfo {pages} {351} (\bibinfo {year} {1998})},\ \Eprint
  {http://arxiv.org/abs/hep-ph/9803298} {arXiv:hep-ph/9803298 [hep-ph]}
  \BibitemShut {NoStop}%
%%CITATION = HEP-PH/9803298;%%
\end{thebibliography}%

\end{document}